\documentclass[11pt]{article}
\input{epsf}
\usepackage{epsfig,amsmath,amssymb}

\def\be{\begin{equation}}
\def\ee{\end{equation}}
\def\ba{\begin{align}}
\def\ea{\end{align}}
\def\beq{\begin{eqnarray}}
\def\eeq{\end{eqnarray}}

\begin{document}

 \title{Bound states in $N=2$ Liouville theory with boundary \\
and 
\\
Deep throat D-branes}

 \author{Raphael Benichou and Jan Troost}
 
\maketitle

\begin{center}
 \emph{Laboratoire de Physique Th\'eorique} \footnote{Unit\'e Mixte du CNRS et
    de l'Ecole Normale Sup\'erieure associ\'ee \`a l'universit\'e Pierre et
    Marie Curie 6, UMR
    8549. Preprint LPTENS-08/29.},
\emph{ Ecole Normale Sup\'erieure,  \\
24 rue Lhomond, F--75231 Paris Cedex 05, France}
\end{center}

 \begin{abstract} 
We exhibit bound states in the spectrum of non-compact D-branes in $N=2$
 Liouville conformal field theory. We interpret these
states in
the study of D-branes in the near-horizon limit of Neveu-Schwarz five-branes spread on a
 topologically trivial circle. 
We match semi-classical 
di-electric and repulsion effects with exact conformal field theory
results and
describe the fate of D-branes hitting NS5-branes.
We also show that the bound states can give rise
 to massless vector and hyper multiplets in a low-energy gauge theory on
 D-branes deep inside the throat.
\end{abstract}

\section{Introduction}
Liouville conformal field theory plays a central role in
two-dimensional gravity and string theory as well as in our present
understanding of conformal field theories with continuous spectrum.
When rendered $N=2$ supersymmetric, it becomes a building block for
supersymmetric string backgrounds. We further study $N=2$ superconformal
Liouville theory with boundary in this paper, and concentrate in particular
on exhibiting bound states on non-compact branes.

In a separate development, an intuitive description of large parts of
gauge theory physics has been given in terms of brane set-ups (see
e.g. \cite{Hanany:1996ie}\cite{Witten:1997sc}\cite{Elitzur:1997fh}). We
believe it is an interesting task to 
study
descriptions of brane set-ups that are
 valid at all energy scales. 
 It further brings to bare string theory techniques in gauge theory
physics and vice versa.  
For instance, one may suspect that
world-sheet $N=2$ holomorphy will turn out to be a useful alternative
tool to analyze space-time physics.

The analysis of D-branes in Neveu-Schwarz five-brane backgrounds
\cite{Elitzur:2000pq} was revisited in \cite{Israel:2005fn} using the
tools developed to solve non-rational conformal field theories. It was
argued how to realize brane set-ups \cite{Hanany:1996ie} in terms of
boundary conformal field theories (see also
\cite{Lerche:2000jb}\cite{Eguchi:2003ik}).

In this paper we wish to analyze in more detail than in
\cite{Israel:2005fn} how a D-brane behaves in the close neighborhood
of NS5-branes. We will study a particular set-up where the NS5-branes
are evenly arranged on a topologically trivial circle.  We will see
how target space physics is coded in intricate properties of $N=2$
Liouville theory with boundary.  In section \ref{cft} we lay bare the
relevant properties of $N=2$ Liouville theory with boundary, which
include new relations between formal boundary states as well as the
appearance of bound states in boundary spectra. In section \ref{bulk}
we review briefly the bulk conformal field theory. We construct
D-branes in the full superstring theory in section \ref{DbraneCFT},
and analyze their spectrum.
In section \ref{D4NS5} we give a space-time interpretation of the
conformal field theory results of section \ref{cft}. In particular we
show the appearance of light bound states on the world-volume of
non-compact D-branes as they approach NS5-branes. We also describe the
cutting of D-branes as they hit NS5-branes.  We analyze the gauge
theory excitations living deep inside the 
triply scaled throat in
section \ref{gauge}. In section \ref{two} we discuss a special case
with more symmetry allowing us to make a crosscheck, 
and we conclude with a summary and suggestions for
further developments in section \ref{conclusions}. Various detailed
calculations are gathered in the appendices for the benefit of the
indefatigable reader.


\section{Addition relations and bound states on the boundary}
\label{cft}
We first
analyze 
aspects of $N=2$ Liouville theory with boundary
(see e.g. \cite{Eguchi:2003ik}\cite{Ahn:2003tt}\cite{Israel:2004jt}\cite{Ahn:2004qb}\cite{Hosomichi:2004ph}).
We will 
derive new addition relations between boundary states, and
show the appearance of normalizable boundary vertex 
operators as bound states on non-compact branes.

\subsection{Unitary representations of the $N=2$ superconformal
  algebra}
We take the central charge of our $N=2$ Liouville model to be
\be c=3 + \frac{6}{k} \ee
where $k$ is a strictly positive integer.
States in $N=2$ Liouville theory can be organized in
representations of the $N=2$ superconformal algebra. The characters
of the 
representations can be obtained as coset characters from the
$N=1$ superconformal extension of an $SL(2,\mathbb{R})$ 
current algebra
\cite{Dixon:1989cg} after gauging a compact $U(1)$ subalgebra. 
In the following we use the $SL(2,\mathbb{R})$ quantum
numbers to label the representations of the $N=2$
superconformal algebra. After descent
in the NS-sector, the primary state with Casimir and
$U(1)$ labels $(J,m)$  has
conformal dimension $h$ and $U(1)_R$
charge $Q$ determined by the formulas:
\be\label{hQ} h= -\frac{J(J-1)}{k}+ \frac{m^2}{k}, \qquad Q = \frac{2m}{k}\ee
We take the R-charge $Q$ 
to be quantized such
that $2m \in \mathbb{Z}$.

We have different families of
unitary representations:
\begin{itemize}
\item Continuous characters
 ($0 \le J \le \frac{1}{2}$ or $J=\frac{1}{2} + i P$, $P
  \in \mathbb{R}^+$; $m$ 
 generic): \be ch_c(J,m;\tau,z) =
  q^{-\frac{J(J-1)}{k}+ \frac{m^2}{k}-\frac{1}{4k}} y^{\frac{2m}{k}}
  \frac{\theta_3(\tau,z)}{\eta^3(\tau)} \ee These representations have no null
  vectors. The parent $SL(2,\mathbb{R})$ representation has a
  spectrum for a compact generator which is doubly infinite.
  
\item Discrete representations 
  ($J=\frac{1}{2},1,\dots,\frac{k+1}{2}$ ; $m=J+r$, $r \in \mathbb{Z}$
): \be ch_{d}(J,r;\tau,z) = q^{-\frac{(J-1/2)^2}{k}+\frac{(J+r)^2}{k}}
y^{\frac{2(J+r)}{k}} \frac{1}{1+yq^{1/2+r}}
\frac{\theta_3(\tau,z)}{\eta^3(\tau)} \ee These have one null vector
and are inherited from a representation of $SL(2,\mathbb{R})$ with
semi-infinite spectrum. When $r$ is negative, the conformal dimension
and R-charge of the primary are given by formulae (\ref{hQ}) with
$\tilde{m}=J+r-k/2$ and $\tilde{J}=(k+2)/2-J$ (see
\cite{Israel:2004jt}).

\item Identity representation ($J=0$; $m \in \mathbb{Z}$):
\be ch_{\mathbb{I}}(m;\tau,z) = q^{-\frac{1}{4k}+\frac{m^2}{k}}
y^{\frac{2m}{k}} \frac{1-q}{(1+y q^{\frac{1}{2}+m})(1+y^{-1}
  q^{\frac{1}{2}-m})} \frac{\theta_3(\tau,z)}{\eta^3(\tau)} \ee
  This representation has two null vectors. It 
descends from the trivial 
representation of the $SL(2,\mathbb{R})$ algebra.
\end{itemize}
 We hope the notations are sufficiently intuitive and refer to
\cite{Dixon:1989cg} and \cite{Israel:2004jt}
for further details.

The chiral and anti-chiral rings of the $N=2$ superconformal field
theory form interesting subsectors of the theory.  The chiral ring is
generated by the operators that satisfy $h=\frac{Q}{2}$. In the
anti-chiral ring, the relation $h=-\frac{Q}{2}$ is satisfied. We find
a chiral primary state in the discrete character when $r=0$, and an
anti-chiral primary state in the discrete character when $r=-1$ (see
\cite{Israel:2004jt}). The identity character with $m=0$ contains the
identity operator ($h=Q=0$) which is both chiral and anti-chiral.

Interpreted as a $\sigma$-model, 
$N=2$ Liouville theory 
has a curved non-compact target space. States in continuous representations
with $J=\frac{1}{2}+iP$ propagate to infinity. The momentum in the
curved 
non-compact direction is denoted by $P$. 
States in discrete or identity representations are bound states.

The characters of the ${N}=2$ superconformal algebra satisfy the
following identity  \cite{Eguchi:2003ik}: 
\be\label{1stId} ch_{c} (J=0,m=0) = ch_{I} (m=0)  +
ch_{d}(J=\frac{k}{2},r=0) + ch_{d} (J=1,r=-1). \ee
as well as the identity 
\be\label{2ndId} ch_{c} (J=\frac{1}{2}, m=\frac{1}{2}) =
ch_{d}(J=\frac{1}{2},r=0) + ch_{d} (J=\frac{k+1}{2},r=0). \ee
These identities can be viewed as 
generalizations of the
identities between discrete, continuous and the trivial representations of
$SL(2,\mathbb{R})$. 
One can write down such an identity for any continuous character, if
there exists a discrete or a 
finite
dimensional representation with the same
quantum numbers. 
These additional addition relations are
either spectral flowed from (\ref{1stId}) and (\ref{2ndId}) 
(see later), or
involve non-unitary characters.

Note that the case $k=1$ is special. In that case, we can combine the two relations (\ref{1stId}) and
(\ref{2ndId}) to obtain an identity involving only the identity and continuous
characters: 
\be \label{k=1Id} ch_{c}^{k=1} (J=0,m=0) = ch_{I}^{k=1} (m=0)  +
ch_{c}^{k=1}(J=\frac{1}{2},m=\frac{1}{2}). \ee

\subsection{Spectral flow and extended characters}

When we simultaneously change boundary conditions on all fermionic operators
of the $N=2$ superconformal algebra (such that they pick up a phase as one
goes around a world-sheet circle), we spectral flow the $N=2$ algebra. After
one unit of spectral flow, we recuperate the original boundary
conditions. This operation 
acts on the quantum numbers $(J,m)$ of a given state 
as $(J,m) \to (J,m+1)$.

We define the extended $ {N}=2$ superconformal algebra as being
generated by the usual $ {N}=2$ generators together with the operator
implementing $k$ units of spectral flow. For applications in superstring
theory and to simplify the modular properties of the characters, it is
convenient 
to classify the states in
representations of the extended algebra. We define the following extended
characters:
\be Ch_{c}(J,m;\tau,z) = \sum_{n \in \mathbb{Z}} ch_c(J,m+kn;\tau,z)
\label{ExtC} 
\ee 
\be Ch_{d}(J,r;\tau,z) = \sum_{n \in \mathbb{Z}}
 ch_d(J,r+kn;\tau,z)\label{ExtD} 
 \ee
\be Ch_{I}(m;\tau,z) = \sum_{n \in \mathbb{Z}} ch_{\mathbb{I}}(m+kn;\tau,z) \label{ExtI} \ee
The quantum numbers $m$ and $r$ 
labeling the extended characters
are
defined modulo $k$. The behavior of
 these extended characters under modular transformations is given in appendix
 \ref{discretecouplings}.
We can perform spectral flow on both sides of the identities (\ref{1stId}) and
(\ref{2ndId}) (by any amount) and obtain new
identities. In particular, the character identities 
(\ref{1stId}) and (\ref{2ndId})
are valid for
the extended characters as well: 
\be \label{1stIdEx} Ch_{c} (J=0,m=r) = Ch_{\mathbb{I}} (m=r)  +
Ch_{d}(J=\frac{k}{2},r) + Ch_{d} (J=1,r-1). \ee
\be \label{2ndIdEx} Ch_{c} (J=\frac{1}{2}, m=\frac{1}{2}+r) =
Ch_{d}(J=\frac{1}{2},r) + Ch_{d} (J=\frac{k+1}{2},r). \ee

\subsection{Boundary states}\label{boundary}

We consider A-type branes in $  {N}=2$ Liouville theory
(see e.g. \cite{Eguchi:2003ik}\cite{Israel:2004jt}\cite{Hosomichi:2004ph}). 
The boundary
state associated to a brane $B$ is written as:
\be\label{Cardy} |B \rangle = \int_{P=0}^{\infty} \sum_{2m=0}^{2k-1} \psi^{B}_c (P,m) |P,m
\rangle \rangle_c + \sum_{2J=2}^{k} \sum_{2m=0}^{2k-1} \psi^{B}_d (J,m) |J,m
\rangle \rangle_d \ee
The definition of the continuous and discrete Ishibashi states $ |P,m \rangle
\rangle_c$ and $|J,m \rangle \rangle_d$, as well as the 
one-point 
couplings to
the discrete operators $\psi^{B}_d (J,m)$, are given in appendix
\ref{discretecouplings}. 
In the bulk of the paper, we focus on the 
one-point functions of the
continuous operators $\psi^{B}_c (P,m)$.

The identity brane $|0 \rangle_{\mathbb{I}}$
 whose  self-overlap is the extended identity 
 character 
$Ch_{\mathbb{I}}(m=0)$ (in the 
boundary channel)
 has the following one-point function: 

\begin{eqnarray}
\Psi^{\mathbb{I}}_0 (P,m) &=&
 \nu^{iP} \frac{\Gamma(\frac{1}{2}+iP+m) \Gamma(\frac{1}{2}+iP-m)}{
\Gamma(1+2iP) \Gamma(\frac{2iP}{k})}.
\end{eqnarray}
The parameter $\nu$ is related 
to the coefficient $\mu$ of the bulk Liouville potential as
$\nu=\mu^{2/k}$ (see  \cite{Ahn:2003tt}\cite{Hosomichi:2004ph}).
The branes whose overlap with the identity brane $|0 \rangle_{\mathbb{I}}$ are 
 one of the extended characters
 (\ref{ExtC}), (\ref{ExtD}) or
(\ref{ExtI}) are specified by
 the one-point functions:
\begin{itemize}
\item Continuous brane $|\hat P, \hat m \rangle_c $ :
 \begin{eqnarray}\label{psiC}
\psi_{\hat P, \hat m}^{c} (P,m) &=& 
\nu^{iP} \frac{2}{k} e^{- 2 \pi i \frac{2m.2 \hat m}{2k}}
\cos (4 \pi P \hat P/k)
\frac{\Gamma(1-2iP) \Gamma(\frac{-2iP}{k})}{
\Gamma(\frac{1}{2}-iP+m) \Gamma(\frac{1}{2}-iP-m)}.
\end{eqnarray}
We can take the definition to apply to boundary states with $\hat P$ purely
imaginary as well. As long as we stay within the bounds $0 \le i \hat P \le \frac{1}{2}$ the
overlap with the identity brane $|0 \rangle_{\mathbb{I}}$ is the continuous character
with real $\hat J=\frac{1}{2}+i \hat P$, $0\le \hat J \le 1/2$, which remains a unitary
representation of the superconformal algebra.

\item Discrete brane $|\hat J, \hat r \rangle_d$ :
\beq
\psi^{d}_{\hat J, \hat r}(P,m) 
&=&  \nu^{iP} \frac{1}{2 k}
e^{-2 \pi i \frac{2(\hat J+
    \hat r). 2m}{ 2k} }
\frac{\Gamma(1-2iP)\Gamma(-2i\frac{P}{k})}{\Gamma(\frac{1}{2}-iP+m)\Gamma(\frac{1}{2}-iP-m)}
\nonumber \\
& & 
\times \frac{\cosh \left( 2 \pi \frac{P (2\hat J-1)}{k} \right) + (-1)^{2m} \cosh
  \left( 2 \pi \frac{P (k-2\hat J+1)}{k} \right) }{| \cosh \pi (P+im) |^2}. \\
\eeq

\item Identity brane $|\hat m \rangle_{\mathbb{I}} $ :
 \be \Psi^{\mathbb{I}}_{\hat m} (P,m) =
 \nu^{iP}  e^{- 2 \pi i \frac{2 m. 2\hat m}{2k}}
\frac{\Gamma(\frac{1}{2}+iP+m) \Gamma(\frac{1}{2}+iP-m)}{
\Gamma(1+2iP) \Gamma(\frac{2iP}{k})}.\ee

\end{itemize}

The Cardy condition has been checked for the continuous and identity
branes. For the discrete branes, this condition is not
satisfied in 
general \cite{Eguchi:2003ik}\cite{Israel:2004jt}\cite{Fotopoulos:2004ut}. 
We show in appendix \ref{discretecouplings} 
that discrete characters generically 
appear
 with non-integer and/or negative multiplicity in
the overlap of two discrete branes. That indicates that the discrete boundary
states are inconsistent.

\subsection{Brane addition relations}

The modular bootstrap approach suggests that the characters identities
(\ref{1stIdEx}) and (\ref{2ndIdEx}) can be extended to the full boundary
states\footnote{Note that if one defines boundary state that corresponds to single
  (non-extended)
 representations
of the $N=2$ superconformal algebra, one also finds analogous
identities between the corresponding boundary states.}.
That is the case for example
in bosonic
Liouville theory (see \cite{Martinec:2003ka}\cite{Teschner:2003qk}). There is
also a known example in supersymmetric Liouville theory at level $k=1$ 
(see \cite{Murthy:2006xt}).
We propose the relations ($\hat{r}\in \mathbb{Z}$):
\beq\label{c=i+d+d} |\hat J=0,\hat r \rangle_c &=& |\hat r
  \rangle_{\mathbb{I}}+ |k/2,\hat r \rangle_d + |1,-1+\hat r
\rangle_d  \eeq
 \beq\label{c=d+d}  |\hat P=0,1/2+\hat r \rangle_c &=& |1/2,\hat
  r \rangle_d +
  |k/2+1/2,\hat r \rangle_d. \eeq
The labels of the brane correspond to the extended character that appears when
computing the annulus partition function 
for boundary operators between the brane
 and the identity brane $|0  \rangle_{\mathbb{I}}$.
After some $\Gamma$-function and trigonometric gymnastics, it can be
shown that the one-point functions for these branes indeed satisfy the
corresponding addition relations. Moreover, the 
identities hold
for the discrete part of
the Cardy states as well.

Since the boundary states for individual discrete branes
do not pass the Cardy check, we cannot view these addition relations as
identities between 
consistent boundary states. Nevertheless, we will see in section \ref{D4NS5}
that these relations code semi-classical properties of
D-branes in string theory.

As a side-remark we note that in the special case $k=1$, 
we get the brane addition relations of
\cite{Murthy:2006xt}:
\begin{eqnarray}
|\hat J=0, \hat r \rangle_c^{k=1} &=& |\hat J=1/2,1/2+\hat r
 \rangle_c^{k=1} + |\hat r
 \rangle_{\mathbb{I}}^{k=1}. 
\end{eqnarray}
In this case, the addition relation is one between boundary states that pass the Cardy check individually.

\subsection{Bound states on non-compact branes}\label{bound}

We will now investigate more closely the 
first addition relation
(\ref{c=i+d+d}). If one computes the self overlap of the right-hand
side of the relation (\ref{c=i+d+d}), one will obtain states in discrete and in the identity
representations\footnote{The detailed calculation is given in appendix \ref{discretecouplings}.}.
These states are bound states, localized on the world-volume of the
brane. Their conformal dimension 
can lie
below the continuum of boundary vertex
operators. In this section we wish to understand the appearance of these bound
states in the self overlap of the continuous brane on the left-hand side of
(\ref{c=i+d+d}). To that end, we will compute the overlap of two continuous branes $
{}_c\langle \hat P_1,\hat m_1|\mathrm{prop}|\hat P_2,\hat m_2 \rangle_c $,
where ``prop'' represents the bulk (cylinder) propagator. Then we will study what happens 
when we continue
$\hat P_1$ and $\hat P_2$ to $ \frac{i}{2}$ (i.e. $\hat J_1$ and $\hat J_2$ to zero). 
This analysis is inspired by similar
analyses in 
bosonic Liouville theory 
\cite{Teschner:2000md}\cite{Teschner:2003qk}, and 
in supersymmetric Liouville theory at level $k=1$ \cite{Murthy:2006xt}.

We start from the definition of the continuous branes (\ref{Cardy}),
(\ref{psiC}). The continuous branes do not couple to discrete bulk operators.
 The overlap of the two branes
gives in the bulk
channel:
\begin{eqnarray}
Z_{1,2} &=& {}_c\langle \hat P_1,\hat m_1|\mathrm{prop}|\hat P_2,\hat m_2 \rangle_c =
\int_0^\infty dP \sum_{2m \in Z_{2k}}
\frac{4}{k^2} e^{ 2 \pi i \frac{2 m (2\hat m_1-2 \hat m_2) }{2k}} \cos \left( \frac{4 \pi P
  \hat P_1}{k}\right) \cos \left( \frac{4 \pi P \hat P_2}{k} \right)
\nonumber \\
& & 
\frac{\cos \pi (-iP +m) \cos \pi (iP+m)}{ \sinh (2 \pi P) \sinh (\frac{2 \pi
    P}{k}) } Ch_{c}^{closed}(P,m)
\end{eqnarray}
 We obtain the partition function in the loop channel via a modular transformation:

\begin{eqnarray}\label{Zopen1}
Z_{1,2} 
&=&
\int_0^\infty dP 
\frac{1}{k} 
\sum_{\epsilon_{1},\epsilon_{2}=\pm 1}
\cosh \left( 2 \pi P \left(1 + 2 i \epsilon_1 \frac{\hat P_1}{k} + 2 i \epsilon_2
\frac{\hat P_2}{k}\right) \right) 
\frac{1}{ \sinh (2 \pi P) \sinh (\frac{2 \pi P}{k})} 
\nonumber \\
& & 
\int_{- \infty}^\infty dP'
e^{\frac{4 \pi i P P'}{k}} Ch^{open}_{c}(P',\hat m_1-\hat m_2)
\nonumber \\
&+& 
 \int_0^\infty dP 
\frac{2}{k} \sum_{\epsilon_2=\pm 1} \cosh \left( 2 \pi P \left( 2 i \frac{\hat P_1}{k} + 2
i \epsilon_2 \frac{\hat P_2}{k} \right) \right)
\frac{1}{ \sinh (2 \pi P) \sinh (\frac{2 \pi P}{k})} 
\nonumber \\
& & 
\int_{-\infty}^\infty dP'
e^{\frac{4 \pi i P P'}{k}} Ch^{open}_{c}\left( P',\hat m_1- \hat
  m_2+\frac{k}{2} \right)
\end{eqnarray}
Now we would like to write the boundary partition function $Z_{1,2}$ in a
more explicit form as
an integral over boundary operators with a given spectral density:
\beq Z_{1,2} &=& \int_{P'=-\infty+i\Delta}^{\infty+i\Delta} 
\rho_1(P';\hat P_1, \hat P_2)
Ch_c^{open}(P',\hat m_1- \hat m_2) \nonumber \\
&&+ \int_{P'=-\infty+i\Delta}^{\infty+i\Delta} 
\rho_2(P';\hat P_1,\hat P_2)
Ch_c^{open}\left( P',\hat m_1-\hat m_2+\frac{k}{2} \right).
 \eeq
 Notice the imaginary shift $\Delta$ in the integration contour. This shift is
 needed for the following reason: in order to make the necessary exchange of
 the bulk and boundary momentum integrals, we need the integral over the bulk
 momentum $P$ at fixed boundary momentum $P'$ to be convergent.  Looking at
 the integrand in (\ref{Zopen1}), we identify two possible divergences at
 $P=0$ and $P \to \infty$. 
Firstly, the integral is always divergent when the bulk
 momentum $P$ goes to zero. 
 Notice that the behavior of the
 integrand close to $P=0$ is universal: it does not depend on the brane
 parameters $\hat P_1$, $\hat P_2$. It is an infrared divergence due to the
 non-compactness of the target space.
We can cure the divergence either by
considering a relative spectral density (subtracting a reference spectral
density
 with fixed value 
of the brane parameters), or by removing the pole
 by hand.
Secondly, the bulk momentum integral may still diverge due to the 
ultraviolet integration region. A prescription to cure this divergence
\cite{Teschner:2000md} is
to lift the
 integration contour for the boundary momentum
$P'$ by giving it  a sufficient imaginary part $\Delta$.
 Then the integrand will decay exponentially at large momentum $P$, and we
 will be able to exchange the order of integration. 
 
 Let's compute the necessary shift of the contour $\Delta$. At large momentum
 $P$, the integrand in the first two lines of equation (\ref{Zopen1}) behaves
 as: $\exp \left( \frac{2\pi P}{k} (2i\epsilon_1 \hat P_1 + 2i\epsilon_2 \hat
   P_2 -1 + 2iP' )\right)$. The integrand in the last two lines of
equation
 (\ref{Zopen1}) is always less divergent. We deduce that we need to shift the
 integration contour when $2 Im(\hat P_1) + 2 Im(\hat P_2) \ge 1$. We recall
 that we took the imaginary parts of $\hat{P}_{1,2}$ to be positive. The
 imaginary part we give to the boundary momentum $P'$ is: 
\be\label{a}
 \Delta=\max \left( 0\, ,\, Im(\hat P_1) + Im(\hat P_2) - \frac{1}{2} + \epsilon \right) \ee with
 $0<\epsilon \ll 1$. Since 
we have the inequality
$0 \le Im(\hat P_1),Im(\hat P_2) \le \frac{1}{2}$,
 we need both $\hat P_1$ and $\hat P_2$ imaginary to have a non-zero shift (or
 one of them equal to $i/2$, in which case $\Delta=\epsilon$).

After giving a sufficiently large imaginary part to the boundary momentum
(which breaks parity symmetry in $P'$), we find the 
following 
densities of boundary operators: 
 \begin{eqnarray}\label{rhos}
 \rho_1(P';\hat P_1,\hat P_2) &=&
 \int_0^\infty dP 
\frac{1}{k} 
\sum_{\epsilon_{1},\epsilon_2=\pm1}
\frac{\cosh \left( 2 \pi P \left(1 + 2 i \epsilon_1 \frac{\hat P_1}{k} + 2 i \epsilon_2
 \frac{\hat P_2}{k}\right) \right)}{ \sinh (2 \pi P) 
\sinh (\frac{2 \pi P}{k})} 
e^{\frac{4 \pi i P P'}{k}} 
\nonumber \\
\rho_2(P';\hat P_1, \hat P_2) &=&
 \int_0^\infty dP 
\frac{2}{k} \sum_{\epsilon_2=\pm1} 
\frac{ \cosh \left(2 \pi P \left( 2 i \frac{\hat P_1}{k} + 2 i \epsilon_2 \frac{\hat
 P_2}{k}\right) \right)}{ \sinh (2 \pi P) \sinh (\frac{2 \pi P}{k})} 
e^{\frac{4 \pi i P P'}{k}} .
 \end{eqnarray} 
 The densities $\rho_1$ and $\rho_2$ are functions of the boundary momentum
 $P'$ 
as well as the two brane parameters $\hat P_1$ and $\hat
 P_2$. 
 For given brane parameters, one can 
extend these functions over the whole
 complex plane in $P'$, and identify their poles. One can
 then bring the $P'$ integral back to the real line (in order for the
 continuous part of the spectrum to correspond to continuous characters), 
pick up the poles crossed
 in the process, and identify them as isolated contributions to the partition
 function.

To analyze the pole structure of the densities of states $\rho_1$ and
$\rho_2$, we can rewrite 
these 
densities
as the derivatives of phase
shifts.
We can perform the rewriting in terms of the q-gamma function
$\Gamma_b(x)$. It is defined as (see e.g. \cite{Ponsot:2001ng}):
\begin{eqnarray}
\log \Gamma_b(x) &=&
\int_0^\infty \frac{dt}{t}
\frac{e^{-xt}-e^{Qt/2}}{(1-e^{-bt})(1-e^{-t/b})}
- \frac{(Q-2x)^2}{8 e^t} - \frac{Q-2x}{t}
\end{eqnarray}
where $Q=b+b^{-1}$. This function has poles at $x = -mb -n b^{-1}$, where $m$
and $n$ are positive integers.
 With $b=\sqrt{k}$ and $t=2 \pi p$,
 we can write $\rho_1$ and $\rho_2$ in terms of  q-gamma functions as :
\beq \rho_1(P';\hat P_1, \hat P_2)&=& \frac{1}{2\pi i}  
\sum_{\epsilon_1,\epsilon_2=\pm 1} \frac{\partial}{\partial P'}  
\left[ \log \Gamma_{\sqrt{k}} \left( \frac{-iP'+\frac{1}{2} + i\epsilon_1 \hat
      P_1
      +  i\epsilon_2 \hat P_2}{\sqrt{k}} \right)
 \right. \nonumber \\
&& \left. + \log \Gamma_{\sqrt{k}} 
\left( \frac{-iP'+\frac{1}{2}+k + i\epsilon_1 \hat P_1 +  i\epsilon_2 \hat P_2}{\sqrt{k}} \right) \right] \eeq
\beq \rho_2(P';\hat P_1, \hat P_2)&=&\frac{2}{2\pi i}  \sum_{\epsilon_1,\epsilon_2=\pm 1}
 \frac{\partial}{\partial P'} \left[ \log \Gamma_{\sqrt{k}} \left(
     \frac{-iP'+\frac{1}{2}+\frac{k}{2} + i\epsilon_1 \hat P_1 +  i\epsilon_2
       \hat P_2}{\sqrt{k}} \right) \right].
  \eeq
We note 
that we  subtracted a universal part in the densities that does not
depend on the parameters of the branes $\hat P_1$ and $\hat P_2$. The subtraction regularizes 
the bulk infrared divergence due to the infinite volume of the target space.

\begin{figure}
\centering
\includegraphics[scale=0.9]{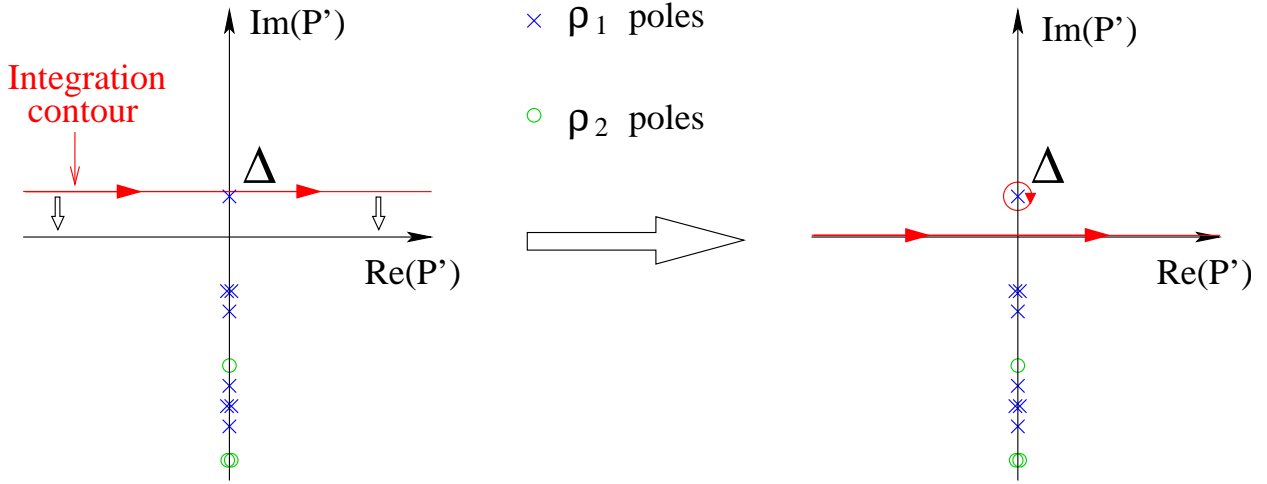}
\caption{ As we shift back to the real axis the integration contour for the
  boundary momentum, we pick up an additional contribution from a pole in the density
  of states $\rho_1$. It gives an isolated contribution to the partition
  function. \label{integrationContour}}
\end{figure}

We can identify which poles are met when we bring back the $P'$
 integration contour
 to the real axis (cf. figure \ref{integrationContour}).
 For the density $\rho_1$, the poles are at:
\begin{eqnarray}
\label{poles} 
-iP'  &=&  -1/2 - n-km + i \epsilon_1 \hat P_1 + i \epsilon_2 \hat P_2
\nonumber \\
-iP' & = &  -1/2 - k - n -km - i \epsilon_1 \hat P_1 - i \epsilon_2 \hat P_2 
\end{eqnarray}
whereas for the density $\rho_2$, we have poles at:
\be 
-iP'  = -1/2 - k/2 - n -km - i \epsilon_1 \hat P_1 - i \epsilon_2 \hat P_2 
\ee
where $\epsilon_1,\ \epsilon_2=\pm1$, and $m$ and $n$ are positive
 integers\footnote{The location of the poles can be determined by
 analyzing the integrand at large bulk momentum $P$.}.
Given that $0 \le Im(\hat P_1),Im(\hat P_2) \le \frac{1}{2}$, there
 is only one pole in 
the spectral density
$\rho_1$ that will be crossed as we shift the boundary momentum 
 contour to the real axis: the one from the first series in 
equation
 (\ref{poles}) with
 $m=n=0$ and $\epsilon_1=\epsilon_2=-1$. The residue for this pole is $+1$.
 The contour of the integral with the spectral
 density $\rho_2$ will not cross a pole when $k$ is greater
than one.

 Eventually we can write down the annulus partition function in an explicit form:
\begin{itemize}
\item When $\Delta = 0 $:
\beq Z_{1,2} &=& \int_{P'=-\infty}^{\infty} dP' \rho_1(P';\hat P_1,\hat P_2)
Ch_c^{open}(P',\hat m_1- \hat m_2) \nonumber \\
&+& \int_{P'=-\infty}^{\infty} dP' \rho_2(P';\hat P_1, \hat P_2)
Ch_c^{open}\left( P',\hat m_1- \hat m_2+\frac{k}{2} \right) \eeq
\item When $\Delta \neq 0 $:
\beq Z_{1,2} &=& \int_{P'=-\infty}^{\infty} dP' \rho_1(P';\hat P_1, \hat P_2)
Ch_c^{open}(P',\hat m_1-\hat m_2) \nonumber \\
&+& \int_{P'=-\infty}^{\infty} dP' \rho_2(P';\hat P_1, \hat P_2)
Ch_c^{open}\left( P',\hat m_1-\hat m_2+\frac{k}{2} \right)  \nonumber \\
& +&  Ch_c^{open}\left( P'=\hat P_1+ \hat P_2-\frac{i}{2},\hat m_1-\hat m_2 \right) \eeq
\end{itemize}

\subsubsection*{Self-overlap of a continuous brane}

Let's examine the self-overlap of a continuous brane with parameters $(\hat
P,\hat m)$. This is just a special case of the previous calculation when we
take
the values
$\hat P=\hat P_1=\hat P_2$ and $\hat m=\hat m_1=\hat m_2$. So we obtain:
\begin{itemize}
\item 
When the brane parameter
$\hat P$ is real, or between $0 \le Im(\hat P)\ < \frac{1}{4}$:
\beq\label{Zc0}  {}_c\langle \hat P,\hat m|\mathrm{prop}|\hat P,\hat m
\rangle_c & = &
\int_{-\infty}^{\infty} dP'   \left( \rho_1(P';\hat P)
Ch_c^{open}(P',0) 
+  
\rho_2(P';\hat P)
Ch_c^{open}\left(P',\frac{k}{2}\right) \right) 
\nonumber \\
& & 
 \eeq
\item 
When the brane parameter is in the range
$\frac{1}{4} < Im(\hat P)\le\frac{1}{2}$:
\beq\label{Zca} {}_c\langle \hat P,\hat m|\mathrm{prop}|\hat P,\hat m \rangle_c &=&
\int_{-\infty}^{\infty} dP' \left( \rho_1(P';\hat P)
Ch_c^{open}(P',0) +  \rho_2(P';\hat P)
Ch_c^{open}\left(P',\frac{k}{2}\right)  \right) \nonumber \\
& + & Ch_c^{open}(J'=1-2Im(\hat P),m=0) \eeq
\end{itemize}
An additional localized contribution appears 
starting
at $Im(\hat P)=\frac{1}{4}$
(namely, at half the momentum intercept caused by the linear dilaton). 
It contains relevant boundary operators, some of which have a conformal dimension below the
continuum.  At 
the value
$Im(\hat P)=\frac{1}{2}$ one of these operators becomes of
conformal dimension zero. There 
 the continuous character $Ch_c^{open}(\hat J=0,\hat m=0)$ splits 
into two discrete and
 one identity characters, according to the identity (\ref{1stIdEx}). If we
 would go further and give $\hat P$ an imaginary part greater than $1/2$, we
 would encounter
 characters
of non-unitary representations. 

We fulfilled our goal to compute the self-overlap of the continuous brane
 appearing on the left-hand side of the brane addition relation
 (\ref{c=i+d+d}):
\beq\label{Z00}
\label{ZzeroJ}
 {}_c\langle \hat J=0,\hat m |\mathrm{prop}|\hat J=0,\hat m \rangle_c  &=&
 \int_{-\infty}^{\infty} dP' \left( \rho_1(P';\hat P,\hat P)
Ch_c^{open}(P',0) \right. \nonumber \\
& & +  \left. \rho_2(P';\hat P,\hat P)
Ch_c^{open}\left(P',\frac{k}{2}\right) \right)  \nonumber \\
& +& Ch_{\mathbb{I}}^{open} (0)  +
Ch_{d}^{open}(\frac{k}{2},0) + Ch_{d}^{open}(1,-1)
\eeq
In appendix \ref{discretecouplings} it is 
shown that one obtains the same result for the self-overlap of the sum of
branes on the right-hand side of equation (\ref{c=i+d+d}).
It is interesting to note that
 the way the correct multiplicity $+1$ for the discrete characters comes about
in the algebraic calculation is via a multiplicity $+2$ from the
 overlap
of identity brane with the discrete branes in the addition relation, and
 multiplicity $-1$ from the 
 self-overlap of the sum of the discrete boundary states. That fact will be
 important later on.

\subsection*{Remarks}

Firstly, as we noted in passing, the case $k=1$ needs to be studied separately.
For $k=1$ and $\hat J=0$ the self-overlap picks up a pole in the 
density
 $\rho_1$ 
as before. 
Moreover,   
a pole in the spectral density $\rho_2$ coincides with the contour
of integration in that case. That is due to a boundary infrared divergence.
The integration can be regularized with a principal value prescription.
Equivalently, we pick up the pole with weight a half (multiplied by the
prefactor of $2$ associated to the density of states $\rho_2$). That leads to
an extra contribution of a single continuous character with $J=1/2=m$ to the
boundary spectrum.  
It implies that both discrete characters appear with multiplicity two in the
self-overlap of the brane (\ref{Z00}).
The same conclusion is reached independently in appendix
\ref{discretecouplings}. There it is shown that there are no discrete
subtractions in this case.
The upshot is an agreement with the discussion in
\cite{Murthy:2006xt} from both ways of calculating the self-overlap.

Secondly, we obtained the localized spectrum for generic level $k$ in two
ways: on the one hand via the analytic continuation of the self-overlap, and
on the other hand using the full identity and discrete boundary states
including their coupling to the discrete Ishibashi states. We believe that
provides a convincing crosscheck for these two procedures. This is important
because it indicates that we should take the couplings of the discrete branes
to discrete Ishibashi states seriously.
It implies in particular that discrete branes do not (generically) pass the
Cardy check.
 
We note that there may be
exchange of discrete states between boundary states despite the fact that no
pole appears in the density of exchanged bulk operators. Whether a discrete
state is exchanged is determined by its coupling to both branes, and when
these are non-zero, the contribution is weighted by the propagator. When one
brane in the overlap is localized, those contributions
 correspond to the formal prescription
to pick up possible poles and their residues in the bulk channel - why this is
the case can be understood from the modular bootstrap (see
e.g.\cite{Israel:2004jt}). That the prescription to pick up poles in the bulk
channel only works when one of the two branes is localized seems related to
the observation \cite{Jego:2006ta} that an analogue of the Verlinde formula in
non-rational conformal field theories works well only in those cases.

 In the
dual boundary channel, it is the contour dictated by the precise 
boundary conditions
that determine whether or not poles in the density of boundary operators
contribute to the annulus amplitude \cite{Teschner:2000md}.

\subsection{Semi-classical analysis of the
  branes}\label{semiclassicsLiouville}
We finalize this section by a few 
remarks on the semi-classic target
space interpretation of the boundary conformal field theories.  The
semi-classical geometry has been described in e.g.
\cite{Fotopoulos:2003vc}\cite{Ribault:2003ss}\cite{Fotopoulos:2005cn}\cite{Israel:2005fn}
in some detail.
We add a few extra remarks to those analyses that will
be useful in the following.

When we think
 of $ {N}=2$ Liouville theory as a two-dimensional $\sigma$-model, the
 associated 
trumpet target-space geometry is given by:
\be\label{trumpet}
\left \{ \begin{array}{l}
ds^2 = \alpha' k \left( dr^2 + \coth^2 r \ d\psi^2  \right)  \\
 e^{2\Phi} = \frac{e^{2\Phi_0}}{\sinh^2r}
\end{array} \right.
 \ee
 The curvature radius of the geometry is $\sqrt{k \alpha'}$. In the large $k$
 limit, we can perform a useful semi-classical analysis.  This
 analysis will break down close to $r=0$, where the geometry is singular
(but the conformal field theory remains well-defined).
The identity branes are located
 in that region.

\subsubsection*{Semi-classical D1-branes in the trumpet}
Studying solutions that extremize the 
Dirac-Born-Infeld action for D1-branes 
in the background (\ref{trumpet}) (see e.g. \cite{Fotopoulos:2003vc}), 
one finds the following two-parameter family of branes:
\be\label{extBraneLSC} \cosh(r) \sin(\psi-\psi_0) = c \ee
If $c>1$, we put $c\equiv \cosh r_0$. In this case the brane world-volume is
limited to the region $r \ge r_0$. On the other hand, if $c \le 1$, the brane
extends over the whole space-time. When we go to radial infinity,
the function $\sin(\psi-\psi_0)$ goes to zero. So the branes described by
the curve (\ref{extBraneLSC}) have two anti-podal legs in the asymptotic region, 
in the angular directions $\psi=\psi_0$ and $\psi=\psi_0+\pi$. 

These D1-branes are the semi-classical incarnations of the continuous
and discrete branes as we will argue.
 We will also match the parameters $c$ and $\psi_0$ with the quantum numbers that label the Liouville branes. 

\subsubsection*{Continuous branes}

\begin{figure}
\centering
\includegraphics[scale=0.6]{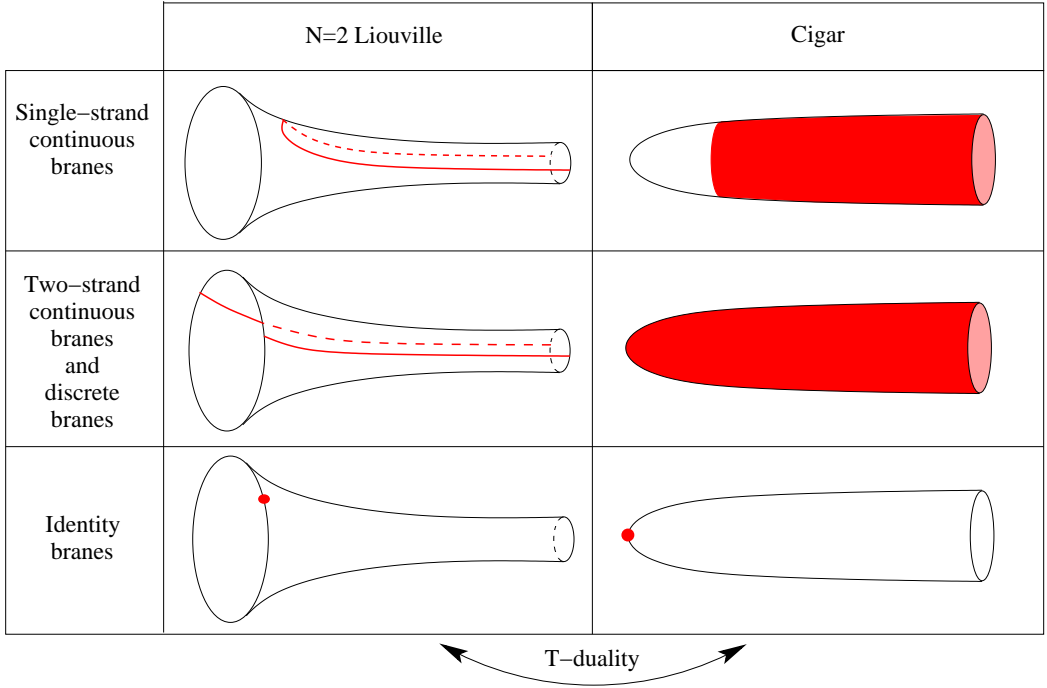}
\caption{Semi-classical description of continuous, discrete and identity
  branes in $N=2$ Liouville theory, and their T-dual image in the cigar.\label{branesLiouville}}
\end{figure}

Let's begin with the continuous branes $|\hat P,\hat m \rangle_c$, with $\hat P$
real and strictly positive. It was argued in \cite{Fotopoulos:2005cn} that the
T-dual D2 brane is doubly sheeted, and partially covers the cigar. We
strengthen the
arguments of \cite{Fotopoulos:2005cn}. First we notice that the parameter
$\psi_0$ defines the position of the brane in the angular direction, so
changing it should not change the self-overlap of the brane\footnote{This
  reasoning only holds far away from the region where the Liouville potential
  is not negligible, since the potential breaks rotation invariance in the
  $\psi$ direction. We can apply it to open strings in continuous
  representations that propagate 
in the asymptotic region.}.  Looking at
equations (\ref{Zc0}) and (\ref{Zca}), we deduce that 
the angle
$\psi_0$ is a function of
the quantum number $\hat m$ only. Moreover the two
 quantities
$\psi_0$ and $\hat m$ are
respectively defined modulo $2\pi$ and $2k$. We are lead to the
identification: 
\be\label{psi0(M)} \psi_0=2\pi \frac{\hat m}{k} - \frac{\pi}{2} \ee 
The additional (harmless) shift is chosen such that the angle $2\pi \frac{\hat
  m}{k}$ 
gives the direction of the tip of the brane (it is also the direction
  of an axis of symmetry for the brane).
Now let's consider the overlap of the brane $|\hat P,\hat m\rangle_c$ with the
identity brane $|\hat m\rangle_{\mathbb{I}}$.
In the 
boundary channel, it is  the continuous extended character
$Ch_c(\hat P,0)$. Thus all the open strings stretching between the two
branes are massive.
We deduce that the brane does not extend up to the horn at $r=0$, and the
parameter $c$ of the semi-classical brane is greater than one. 
To compute the relation between the
parameters $c$ and $\hat P$, we
demand that the mass of a string be equal to its length times its
tension. Equation (\ref{hQ}) implies that the mass square of the lightest open
string is $P^2/k\alpha'$. On the other hand, a string stretching from
the identity brane at $r=0$  up to
the tip of the continuous brane at $r=r_0$ has proper length
$r_0\sqrt{k\alpha'}$, and mass 
$r_0\sqrt{k}/2\pi\sqrt{\alpha'}$.
 Since $c=\cosh r_0$, we deduce that in the semi-classical limit:
\be\label{c(P)} c = \cosh \left( 2\pi \frac{\hat P}{k} \right) \ee
In the case where the brane label
$\hat P$ is imaginary (and $\hat J$ is real), we expect
 the analytic continuation of 
the matching relation (\ref{c(P)}) to be valid:
\be c = \cos \left( 2\pi \frac{2\hat J-1}{2k} \right). \ee
In this case the brane extends up to the horn of the trumpet, and is made  of
two disconnected strands (see figure \ref{branesLiouville}).

\subsubsection*{Discrete branes}
The discrete branes have couplings to the continuous
representations which are identical to those of the D2-branes covering the
whole cigar
\cite{Ribault:2003ss}. This  can be seen by rewriting:
\begin{eqnarray}
\psi^{d}_{\hat J,\hat r}(P,m) & = & \frac{1}{2 \pi} \nu^{iP}  e^{-2 \pi i
  \frac{2m.2\hat m}{2k}} \Gamma(-2iP)\Gamma(1-2i\frac{P}{k}) 
\nonumber \\
& &  \left( \frac{\Gamma(\frac{1}{2}+iP+m)}{\Gamma(\frac{1}{2}-iP+m)}
e^{i \pi ( \frac{1}{2} +\frac{1-2\hat J}{k} ) (2iP)}
+ \frac{\Gamma(\frac{1}{2}+iP-m)}{\Gamma(\frac{1}{2}-iP-m)}
e^{-i \pi ( \frac{1}{2} +\frac{1-2\hat J}{k} ) (2iP)} \right).
\label{onepointdiscrete}
\end{eqnarray} 
We identify the one-point functions after a T-duality transformation on the parameters.
As for the continuous brane, the angular parameter $\psi_0$ is given by 
the quantum number $\hat m$:
\be\label{psi0(r)d}
\psi_0=2\pi \frac{\hat m}{k} - \frac{\pi}{2} \ee To precisely match the parameter $c$ of
the semi-classical brane with the quantum numbers $\hat J$,$\hat r$ defining
the discrete brane, we translate the 
semi-classical matching of the spectral 
density with the derivative of the reflection amplitude
of \cite{Ribault:2003ss} into our
T-dual picture. We again obtain: 
\be\label{c(J)d} c = \cos \left( 2\pi
  \frac{2\hat J-1}{2k} \right). \ee

\subsubsection*{The number of branches}

We note that the self-overlap of the discrete brane contains 
continuous representations with angular momentum
$m=0$ and 
$m=k/2$ (cf. appendix \ref{discretecouplings}). Since
 the discrete branes have Dirichlet boundary conditions in the angular
 direction, we conclude (after appropriately normalizing)
that open strings living on this brane have integer
 or half-integer winding number. This proves that the discrete branes have
 two anti-podal branches.

We may check this argument directly on the one-point function (\ref{onepointdiscrete}). We take the
infrared limit by concentrating on the behavior of the one-point function for
bulk operators
with nearly zero momentum $P$. An analysis of the poles of the $\Gamma$-functions
shows that the dominant couplings are to even momentum modes, which says that
asymptotically we have two branches (that effectively halve the radius of the
asymptotic circle, thus leading to only even momentum couplings).

The same arguments apply to the continuous branes which therefore also have
two branches.


\section{Neveu-Schwarz five-branes spread on a circle}
\label{bulk}

We now want to use what we have learned about  $N=2$ Liouville theory with
boundary to study the behavior of D-branes close to NS5-branes. Our strategy
will be to study the branes first in the T-dual coset
conformal
field theories, then to 
translate back the results into the NS5-brane background \cite{Israel:2005fn}.
To that end, we briefly review 
the bulk string theory background and its
 T-duals \cite{Ooguri:1995wj}\cite{Kutasov:1995te}.

\subsection{The supergravity solution}

Consider $k$ NS5-branes in type II string theory in flat space and stretching
in the directions with coordinates $x^{\mu=0,1,\dots,5}$. They preserve
sixteen supersymmetries. The coordinates of the transverse space are
$x^{i=6,7,8,9}$. The back-reaction of the NS5-branes on flat space is coded in
the string frame metric, the NSNS three-form $H_{(3)}$ and the dilaton $\Phi$:
\be\label{NS5background}
\left \{ \begin{array}{rcl}
ds^2 &=& \eta_{\mu\nu} dx^{\mu} dx^{\nu} + H \delta_{ij} dx^i dx^j  \\
e^{2 \Phi}  &=& g_s^2 \, H  \\
H_{(3)} &=& \ast_{4} d H
\end{array} \right.
\ee
where the harmonic function $H$ is specified in terms of the positions
$x^{i}_a$ of the NS5-branes in transverse space:
\begin{eqnarray}
H(x^i) &=& 1 + \sum_{a=1}^k \frac{\alpha'}{|x^i-x^i_a|^2},
\end{eqnarray}
and $\ast_4$ denotes the Hodge star operation in the four Euclidean transverse
directions (with flat metric).

\begin{figure}
\centering
\includegraphics{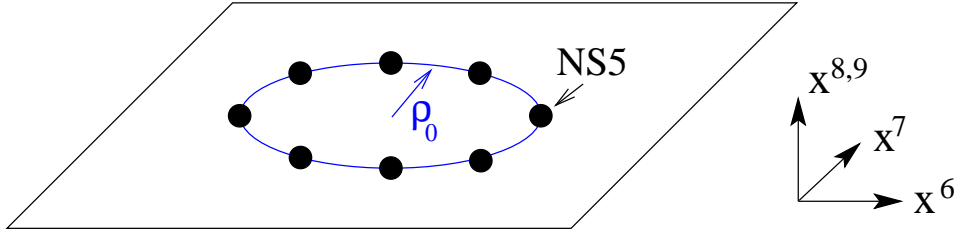}
\caption{We consider $k$ NS5-branes extended in the directions
  $x^{0,1,..,5}$. They are evenly arranged on a circle of radius $\rho_0$, in
  the plane $x^8=x^9=0$. \label{NS5circle}}
\end{figure}

We concentrate on $k$ NS5-branes whose positions are $k$ evenly spread points 
on a topologically trivial circle of radius $\rho_0$ in the $(x^6,x^7)$
plane 
(see figure \ref{NS5circle}). 
We introduce new coordinates $(r,\theta, \psi,\phi)$:
\begin{eqnarray}
(x^6,x^7) &=& \rho_0 \cosh r \sin \theta \, (\cos \psi,\sin \psi)  \nonumber \\
(x^8,x^9) &=& \rho_0 \sinh r \cos \theta \, (\cos \phi,\sin \phi)\,.
\label{cartcoords}
\end{eqnarray}
where $r \in \mathbb{R}_+$, $0\le \theta \le \pi/2$, and $0\le \phi, \psi \le 2\pi$.\\
When we neglect world-sheet instanton corrections (or equivalently, spread the 
NS5-branes over the circle \cite{Sfetsos:1998xd}), and take the double scaling
limit \cite{Giveon:1999px}:
\begin{eqnarray}
g_s, \frac{\rho_0}{\sqrt{\alpha '}} & \to & 0 \nonumber \\
\frac{\rho_0}{g_s \sqrt{\alpha'}}  & & \mbox{fixed},
\end{eqnarray}
the supergravity solution becomes: 
\be\label{tdualNS5geom} \left\{
\begin{array}{rcl}
ds^2 &=& dx^{\mu} dx_{\mu} + \alpha' k \,
\left[ dr^2+d \theta^2 + \frac{\tanh^2 r \ d \phi^2 + \tan^2 \theta \ d \psi^2
}{1+\tan^2 \theta \tanh^2 r} \right] \, , \\
e^{ 2 \Phi} &=& \frac{g_{eff}^2 }{\cosh^2 r - \sin^2 \theta} \, , \\
B  &=& \frac{\alpha' k }{1+\tan^2 \theta \tanh^2 r} \ d \phi \wedge d \psi \,,
\end{array} \right.
\ee
where the effective string coupling constant is
\begin{equation} g_{eff} = 
\frac{\sqrt{k \alpha'}g_s}{\rho_0}. \end{equation}
This background is an
 exact coset conformal field theory, corresponding to
a null gauging of $SU(2) \times SL(2,\mathbb{R})$ \cite{Israel:2004ir}.
However it will be more convenient for us to use the exact conformal field theory description of a
T-dual background.
The
 exact conformal field theory encodes all world-sheet instanton
 corrections \cite{Gregory:1997te}\cite{Tong:2002rq}\cite{Israel:2005fn}.

\subsection{T-dual space-times and their conformal field theory description} 
We perform 
a T-duality along the angular direction $\phi$ and obtain the geometry:
\be \label{geoCFT} \left\{ 
\begin{array}{rcl}
ds^2 &=& dx^{\mu} dx_{\mu} + \alpha' k \left[ d r^2 + \coth^2 r \, \left( \frac{d\omega}{k} + d\psi \right)^2 + d\theta^2 + \mathrm{tan}^2 \theta \, 
\left( \frac{d\omega}{k} \right)^2 \right],  \\
e^{2 \Phi}&=& \frac{g_{eff}^2}{k} \ \frac{1}{\sinh^2 r \cos^2 \theta}
\end{array} \right.
\ee
where $\omega$ is the coordinate T-dual to $\phi$.
The bracketed part of this solution is a $\mathbb{Z}_k$ orbifold of the direct
product of 
two known sigma-models, namely the trumpet (\ref{trumpet}), and the bell defined as:
\be\label{bell} \left\{ \begin{array}{rcl}
ds^2 &=& \alpha' k \left[ d\theta^2 + \tan^2 \theta \, d\omega^2 \right],  \\
e^{2 \Phi}&=& \frac{g_\textsc{eff}^2}{\cos^2 \theta}.
\end{array} \right.
\ee
The latter geometry is the target-space of the  
supercoset $\frac{SU(2)_k}{U(1)}$, also known as the
$N=2$ minimal model of central charge $c=3-6/k$ (see  \cite{Maldacena:2001ky}
 for a review).
The trumpet geometry is more accurately described as $N=2$ Liouville theory
with a momentum condensate.

In order to study the behavior of D-branes in the 
NS5-brane background (\ref{NS5background}), we will use the T-dual description (\ref{geoCFT}).
The conformal field theory describing the curved part of the background is the
$\mathbb{Z}_k$ orbifold of the product
$\left( \frac{SU(2)_k}{U(1)} \right) \, \times
 ( {N}=2\ \mathrm{Liouville})$ . The orbifold 
is the part of the GSO projection that renders the closed string $U(1)_R$
charges
integer.
 To get a full
description of the
 superstring background, we also have to take into account the six flat
directions parallel to the NS5-brane world-volume. They are described by the
$ {N}=2$ conformal field theory of six free real bosons $X^{\mu=0,..,5}$ and six
free real fermions $\psi^{\mu=0,..,5}$. We will work in 
light-cone gauge.
In summary, the relevant critical conformal field theory describing the
dynamics of 
strings in the background (\ref{geoCFT}) is:
\be\label{fullCFT}
\left( \frac{\left(X^{\mu=2,..,5},\psi^{\mu=2,..,5} \right)\times
   \left(\frac{SU(2)_k}{U(1)} \right) 
\times \left( {N}=2 \, \,  \mathrm{Liouville} \right)}{\mathbb{Z}_k^{GSO}}
\right)/ \mathbb{Z}_2^{GSO}
 \ee
The product conformal field theory defines a particular non-compact Gepner
model (see e.g. 
\cite{Eguchi:2000tc}\cite{Eguchi:2004ik}\cite{Ashok:2007ui} for discussions of these models). 

We can also perform T-duality along the isometric $\psi$ direction in the
background (\ref{NS5background}) \cite{Israel:2005fn}. 
That would lead to the background described
by the mirror dual of the conformal field theory (\ref{fullCFT}). In this
mirror description, the $ {N}=2$ Liouville theory is replaced by the cigar
\cite{Eguchi:2004ik}\cite{Ashok:2007ui}.


\section{D-branes in the 
coset conformal field theory}
\label{DbraneCFT}

We will study D-branes in the full superstring theory by first constructing
them in the 
product conformal field theory (\ref{fullCFT}). 
 In section \ref{D4NS5} we re-interpret
them in the Neveu-Schwarz five-brane background. We 
find more explicit realizations and important refinements of the
proposals of \cite{Israel:2005fn}.

\subsection{D-branes in the minimal model and in flat space}\label{MMflat}

The boundary states in the full conformal field theory (\ref{fullCFT}) are the
product of factors coming from Liouville theory, the minimal model and the
free theory. We studied in detail
the boundary states in Liouville theory in section \ref{cft}. Here we
 review briefly the 
relevant boundary states in the minimal model and in flat space.

\subsubsection*{D-branes in flat space}
D-branes in flat space are well-known. They induce Neumann or Dirichlet
boundary conditions for the open strings, according to the directions in which
they are extended. Let's isolate two flat directions in space-time. They are
described on the world-sheet by a free complex boson and 
fermion. It is convenient to bosonize the free fermion, to get a free compact
boson at fermionic radius $\sqrt{2 \alpha'}$. The momentum for this compact boson is given by a
$\mathbb{Z}_4$ integer $s$. 
If this quantum number is even, we are in the Neveu-Schwarz sector while if it is odd,
we are in the Ramond sector.
A D-brane extended in those two directions is
labeled 
by a 
quantum number $\hat{s}$, and its one-point function is:
\be \psi^{N}_{\hat{s}}(s)=\frac{1}{\sqrt{2}}e^{-i\pi \frac{s \bar{s}}{2}}. \ee
For a brane localized in those two directions, the one-point function is:
\be \psi^{D}_{\hat{s},\hat x^\mu}(s,k_\mu)=\frac{1}{\sqrt{2}}
e^{-i\pi \frac{s \bar{s}}{2}} e^{ik_\mu \hat x^\mu}. \ee
In the previous formula, $k_\mu$ is the closed string momentum and $\hat
x^\mu$ gives the position of the brane.

There is one subtlety when we work in the light-cone gauge: we have to put
Dirichlet boundary conditions in the light-cone directions
\cite{Green:1996um}. By a double 
Wick rotation, we can obtain branes extended in
the time direction. The brane will be extended in at most four out of
the six flat directions of (\ref{geoCFT}). In the following we will
consider for definiteness D-branes that fill 
four flat directions.

\subsubsection*{D-branes in the minimal model}

D-branes in the $N=2$ minimal model have also been extensively studied (see
e.g. \cite{Maldacena:2001ky}). A state in this theory is labeled by three
quantum numbers:
an $SU(2)_{k-2}$ spin $j=0,\frac{1}{2},...,\frac{k-2}{2}$, an
angular momentum $n \in \mathbb{Z}_{2k}$, and 
a fermionic number $s \in
\mathbb{Z}_4$. The D-branes with $A$-type boundary conditions
are given by the one-point functions: \be \psi_{\hat j, \hat n, \hat s}^{MM}
(j,n,s) = \frac{1}{\sqrt{k}} \frac{\sin \left( \pi \frac{(2j+1)(2\hat j
      +1)}{k} \right) }{\sqrt{ \sin \pi \frac{2j+1}{k}}} e^{i\pi\frac{n \hat
    n}{k}} e^{-i\pi\frac{s \hat s}{2}}. \ee
The overlap of two branes is given in the open string channel by:
\be \langle \hat j, \hat n, \hat s |\mathrm{prop} | \hat j', \hat n', \hat s'
\rangle  
=
 \sum_j N_{\hat j \hat j'}^j \chi^{open}(j,\hat n-\hat n',\hat s - \hat s') \ee
where $\chi(j, n, s)$ are $N=2$ minimal model characters, 
and $N_{\hat j \hat j'}^j$ are the
 $SU(2)_{k-2}$ fusion coefficients. 
These branes have a clear semi-classical description in the bell geometry
(\ref{bell}), which is weakly curved at large $k$. They are described by the 
equation:
\be\label{braneMM} \sin(\theta) \sin(\omega-\omega_0) = \sin(\theta_0).
 \ee
The geometric parameters $\theta_0$ and $\omega_0$ are related to the brane quantum numbers
$\hat j$ and $\hat n$ 
as \cite{Maldacena:2001ky}:
\be
\label{paramMM} 
\theta_0 = \frac{\pi}{2} - \frac{\pi(2 \hat j + 1)}{k}  \ ,
\quad  \omega_0 = \pi \frac{\hat n}{k} - \frac{\pi}{2} 
\ee
The branes are oriented straight lines connecting 
two points of the boundary of the
 disc geometry.

\subsection{Boundary states in the GSO-projected theory}

We are now ready to write down the boundary states in the full conformal field
theory (\ref{fullCFT}). The last necessary step is to take care of the GSO
projection. As was done in \cite{Ashok:2007ui} for the bulk theories, 
we adapt the standard technology
that exists for compact Gepner models \cite{Recknagel:1997sb} to the
non-compact Gepner models. To
project out the closed string states that do not have odd R-charge, it is
sufficient to project out the Ishibashi states that do not satisfy this
condition. This operation comes with a necessary change of
normalization of the
one-point functions. 
Next, to implement  spin-statistics in the partition
function, we 
add to the one point function a factor $(-1)^{s_0^2/2}$, where $s_0$ is the
fermionic quantum number in one of the flat factors.
That makes for a minus sign for
fermions 
in the one-loop amplitude. Then we can identify the branes that
preserve supersymmetry: they only have states with odd R-charges in their
overlap, in the open string channel. 
This procedure extends to the construction of D-branes in all the non-compact
Gepner 
models studied in \cite{Eguchi:2004ik}\cite{Ashok:2007ui} . 

With these remarks in mind, we can write down the one point functions for the
D-branes. We introduce 
the fermionic numbers $s_{0,1,2,3}$ corresponding respectively 
to
the four factors of the conformal field theory (\ref{fullCFT}): the $X^{\mu=2,3}$
and $X^{\mu=4,5}$ directions, the minimal model and the $N=2$ Liouville factor. 
We distinguish three families of branes corresponding to the three
families 
of section \ref{cft}:
\begin{itemize}
\item Continuous brane $| \hat s_0 ; \hat s_1 ; \hat j, \hat n, \hat
  s_2 ; \hat J,\hat m,\hat s_3 \rangle_c$ :  
\be\label{cBrane} \Psi_{\hat s_0 ;\hat s_1 ; \hat j, \hat n, \hat s_2 ; \hat J,\hat m, \hat s_3}^c = 
\kappa_c e^{i\pi \frac{s_0^2}{2}}
\psi^N_{\hat s_0}
\times \psi^N_{\hat s_1}
\times \psi_{\hat j, \hat n, \hat s}^{MM}
\times 
\psi_{\hat J,\hat m,\hat{s_3}}^{c}
 \ee

\item Discrete brane $| \hat s_0 ; \hat s_1 ; \hat j, \hat n, \hat s_2 ; \hat
  J,\hat r, \hat s_3 \rangle_d$ : 
\be \Psi_{\hat s_0 ; \hat s_1 ; \hat j, \hat n, \hat s_2 ; \hat J, \hat r,
  \hat s_3}^d =  
\kappa_d e^{i\pi \frac{s_0^2}{2}}
\psi^N_{\hat s_0}
\times \psi^N_{\hat s_1}
\times \psi_{\hat j, \hat n, \hat s_2}^{MM}
\times 
\psi_{\hat J,\hat r,\hat{s_3}}^{d}
 \ee
 
\item Identity brane $| \hat s_0 ; \hat s_1 ; \hat j, \hat n, \hat s_2 ; \hat
  m, \hat s_3  \rangle_{\mathbb{I}}$ : 
\be\label{IBrane} \Psi_{\hat s_0 ; \hat s_1 ; \hat j, \hat n, \hat s_2 ; \hat m, \hat
  s_3}^{\mathbb{I}} = 
\kappa_{\mathbb{I}} e^{i\pi \frac{s_0^2}{2}}
\psi^N_{\hat s_0}
\times \psi^N_{\hat s_1}
\times\psi_{\hat j, \hat n, \hat s_2}^{MM}
\times 
\psi_{\hat m,\hat s_3}^{\mathbb{I}}
 \ee
\end{itemize}

The arguments of the one-point functions are left implicit. The additional
 label
 $\hat s_3$ in the Liouville one-point functions $\psi^{c,d,\mathbb{I}}$ 
indicates
 that we consider the sector of the theory 
spectrally flowed by 
$\frac{\hat s_3}{2}$
 units. 
The  normalization factors $\kappa_{c,d,\mathbb{I}}$ are a consequence of the GSO
  projection. They will be fixed by the Cardy condition in the following 
subsection.

\subsubsection*{Annulus amplitudes}

We compute the annulus amplitudes for open strings stretching between the
various families of branes. 
We focus here on
continuous and identity branes. We write out the annulus partition function but suppress the
integral over the modulus of the annulus and the momenta 
(as well as the appropriate measure factor) 
in order not to clutter the formulas.
Let's start out with the overlap of two identity branes \cite{Eguchi:2004ik}\cite{Israel:2004jt}:
\beq\label{Z_II} Z_{\mathbb{I}-\mathbb{I}} &=& {}_{\mathbb{I}} \langle \hat
s_0' ; \hat s_1' ; \hat j', \hat n', \hat s_2' ; \hat m', \hat s_3' |
\mathrm{prop} | \hat s_0 ; \hat s_1 ; \hat j, \hat n, \hat s_2 ; \hat m, \hat
s_3 \rangle_{\mathbb{I}}
\nonumber \\
&=& \kappa_{\mathbb{I}}^2 \frac{1}{K} \sum_{v=0}^{K-1} \frac{1}{2^3}
\sum_{v_1,v_2,v_3=0}^{1} \left[ \left( \sum_{s_0,s_1}
    (-1)^{s_0+\hat s_0 - \hat s_0'} \right. \right. \nonumber \\
& & \left. \delta^{(4)}_{(s_0,2+\hat s_0 - \hat s_0'-v-2v_1-2v_2-2v_3)}
  \delta^{(4)}_{(s_1,\hat s_1 - \hat s_1'-v-2v_1)}
  \frac{\Theta_{s_0,2}}{\eta^3} \frac{\Theta_{s_1,2}}{\eta^3} \right)
\nonumber \\
&& \left( \sum_{j,n,s_2} N_{\hat j \hat j'}^j \delta^{(2k)}_{(n,\hat n- \hat
    n' - v)} \delta^{(4)}_{(s_2,\hat s_2 - \hat s_2'-v-2v_2)} \chi(j, n, s_2)
\right)
\nonumber \\
&& \left. \left( \sum_{r,s_3} \delta^{(2k)}_{(r,\hat m- \hat m' - \frac{v}{2})}
    \delta^{(4)}_{s_3,\hat s_3 - \hat s_3'-v-2v_3} Ch_{\mathbb{I}}(r,s_3)
  \right) \right] \eeq 
We introduced the number $K=lcm(4,2k)$ (namely the
lowest common multiple of $4$ and $2k$).  The summation variables $v$, $v_1$,
$v_2$ and $v_3$ are Lagrange multipliers that implement the GSO projection in
the closed string channel. The summation over $v$ projects the closed string
R-charges to odd integer values. The other three take care of the coherence of
the spin structures between the different factors.
The factor $(-1)^{s_0+\hat s_0 - \hat s_0'}$ gives
a minus sign to the fermionic states, as required by spin-statistics.
 The second line gives the
contribution of the free fermions (at level $2$)
and bosons from the flat directions.
If the two branes are separated by a distance $\delta$ along some flat
direction, there is an 
additional factor $q^{\frac{1}{\alpha'} \left( \frac{\delta}{2\pi} \right)^2}$ in
the partition function. 
Using the Cardy condition, we can deduce the normalization of the identity
brane: 
\be \kappa_{\mathbb{I}}=\sqrt{8K} \ee
We can rewrite this partition function in a more explicit form. 
For the sake of symmetry, we introduce another summation variable $v_0$ equal
to $1-v_1-v_2-v_3$ modulo 2, together with a Lagrange multiplier $a=0,1$ that
enforces this definition. Then we split the summation over $v \in
\mathbb{Z}_K$ into two summations over $p \in \mathbb{Z}_k$ and $b=0,1$.
Finally we perform the summation over the
variables 
$s_0,s_1,j,n,s_2,r,s_3$ to get rid of the delta functions. We obtain:
\begin{eqnarray}
Z_{\mathbb{I}-\mathbb{I}} &=& 
\sum_{v_{0,1,2,3} =0}^{1}  \sum_{a,b=0}^{1} (-1)^b \frac{(-1)^{a(1+\sum v_i)}}{2}
\frac{\Theta_{b+2 v_0 + \hat{s}_0'-\hat{s}_0,2}}{\eta^3} 
\frac{\Theta_{b+2 v_1 + \hat{s}_1'-\hat{s}_1,2}}{\eta^3}
\nonumber \\
& & 
\sum_{2j=0}^{k-2} \sum_{p =0}^{k-1} N^j_{\hat{j} \hat{j}'}
\chi(j,2p+b+\hat{n}-\hat{n}',b+2v_2+\hat{s}_2'-\hat{s}_2)\,
\nonumber \\
& &
Ch_{\mathbb{I}}(p+\frac{b}{2}+\hat m -\hat m',b+2 v_3 + \hat{s}_3' -\hat{s}_3).
\end{eqnarray}
The sum over $v_i$
implements the sum over all possible NS and R-sector states,
such that they do not mix. Since we have the equivalent of four
complex fermions, this is a sum over $\mathbb{Z}_2^4$. 
The sum $\sum v_i$ we can interpret as the fermion number. So the sum over $a$
with the factor of $1/2$ is a projector onto odd fermion
number: $(1-(-1)^F)/2$.
The label $b$ indicates whether we are in the
NS or the R sector, and that the factor $(-1)^b$ implements the space-time
statistics.

%
It is straightforward to generalize the above calculation to the overlap of an
identity brane with a 
continuous brane. We find the same normalization factor for the continuous brane:
 $\kappa_c=\sqrt{8K}$. 
In the open string channel
the overlap reads:
\beq\label{Z_Ic} 
Z_{\mathbb{I}-c} &=&  {}_{\mathbb{I}} \langle \hat s_0' ; \hat s_1' ;
  \hat j', \hat n', \hat s_2' ; \hat m', \hat s_3' | \mathrm{prop} | \hat s_0 ;
  \hat s_1 ; \hat j, \hat n, \hat s_2 ; \hat J, \hat  m, \hat s_3
  \rangle_{c}
\nonumber \\
&=&
\sum_{v=0}^{K-1} 
\sum_{v_1,v_2,v_3=0}^{1} \left[ \left( \sum_{s_0,s_1}
 (-1)^{s_0+\hat s_0 - \hat s_0'} \right. \right. \nonumber \\
& & 
\left. \delta^{(4)}_{(s_0,2+\hat s_0 - \hat s_0'-v-2v_1-2v_2-2v_3)} \delta^{(4)}_{(s_1,\hat s_1 - \hat s_1'-v-2v_1)}
\frac{\Theta_{s_0,2}}{\eta^3} 
\frac{\Theta_{s_1,2}}{\eta^3} \right)
\nonumber \\
&&
\left( \sum_{j,n,s_2} N_{\hat j \hat j'}^j \delta^{(2k)}_{(n,\hat n- \hat n' - v)}
\delta^{(4)}_{(s_2,\hat s_2 - \hat s_2'-v-2v_2)} \chi(j, n, s_2) \right)
\nonumber \\
&&
\left. \left( \sum_{r,s_3} \delta^{(2k)}_{(r,\hat m- \hat m' - \frac{v}{2})} \delta^{(4)}_{s_3,\hat s_3 - \hat
  s_3'-v-2v_3} Ch_{c}(\hat J,r,s_3) \right) \right]
\eeq
Performing the same manipulation of variables as previously, we can rewrite
this partition function as:
\begin{eqnarray}
Z_{\mathbb{I}-c} &=&
\sum_{v_{0,1,2,3} =0}^{1}  \sum_{a,b=0}^{1} (-1)^b \frac{(-1)^{a(1+\sum v_i)}}{2}
\frac{\Theta_{b+2 v_0 + \hat{s}_0'-\hat{s}_0,2}}{\eta^3} 
\frac{\Theta_{b+2 v_1 + \hat{s}_1'-\hat{s}_1,2}}{\eta^3}
\nonumber \\
& & 
\sum_{2j=0}^{k-2} \sum_{p =0}^{k-1} \left[ N^j_{\hat{j} \hat{j}'}
\chi(j,2p+b+\hat{n}-\hat{n}',b+2v_2+\hat{s}_2'-\hat{s}_2) \right. \nonumber \\
&&
 \times \left. Ch_{c}(\hat J, p+\frac{b}{2}+\hat m -\hat m',b+2 v_3 +
   \hat{s}_3' -\hat{s}_3) \right].
\end{eqnarray}
%
  The overlap of two continuous branes in Liouville was computed in section
  \ref{bound}. We deduce the overlap of two continuous branes
  in the product conformal field theory (\ref{fullCFT}), depending on the
  value of the shift
  $\Delta$ defined in equation (\ref{a}). We obtain 
after simplification:

If $\Delta=0$:
\beq\label{Z_cc0} Z_{c-c} &=& 
\sum_{v_{0,1,2,3} =0}^{1}  \sum_{a,b=0}^{1} (-1)^b \frac{(-1)^{a(1+\sum v_i)}}{2}
\frac{\Theta_{b+2 v_0 + \hat{s}_0'-\hat{s}_0,2}}{\eta^3} 
\frac{\Theta_{b+2 v_1 + \hat{s}_1'-\hat{s}_1,2}}{\eta^3}
\nonumber \\
& & 
\sum_{2j=0}^{k-2} \sum_{p =0}^{k-1} \left[ N^j_{\hat{j} \hat{j}'}
\chi(j,2p+b+\hat{n}-\hat{n}',b+2v_2+\hat{s}_2'-\hat{s}_2) \right.
\nonumber \\
& & 
\int_{P'=-\infty}^{\infty} dP' \left( \rho_1(P';\hat P,\hat P')
Ch_c^{open}(P', p+\frac{b}{2}+\hat m -\hat m',b+2 v_3 + \hat{s}_3' -\hat{s}_3)
\right.
\nonumber \\
& &
\ + \left. \left. \rho_2(P';\hat P,\hat P')
Ch_c^{open}(P', p+\frac{b}{2}+\hat m -\hat m',b+2 v_3 + \hat{s}_3' -\hat{s}_3)
\right) \right]
\eeq

If $\Delta \neq 0$:
\beq\label{Z_cca} Z_{c-c} &=& 
\sum_{v_{0,1,2,3} =0}^{1}  \sum_{a,b=0}^{1} (-1)^b \frac{(-1)^{a(1+\sum v_i)}}{2}
\frac{\Theta_{b+2 v_0 + \hat{s}_0'-\hat{s}_0,2}}{\eta^3} 
\frac{\Theta_{b+2 v_1 + \hat{s}_1'-\hat{s}_1,2}}{\eta^3}
\nonumber \\
& & 
\sum_{2j=0}^{k-2} \sum_{p =0}^{k-1} \left[ N^j_{\hat{j} \hat{j}'}
\chi(j,2p+b+\hat{n}-\hat{n}',b+2v_2+\hat{s}_2'-\hat{s}_2) \right.
\nonumber \\
& & 
\left[ \int_{P'=-\infty}^{\infty} dP' \left( \rho_1(P';\hat P,\hat P')
Ch_c^{open}(P', p+\frac{b}{2}+\hat m -\hat m',b+2 v_3 + \hat{s}_3' -\hat{s}_3)
\right. \right.
\nonumber \\
& & 
\ + \left. \rho_2(P';\hat P,\hat P')
Ch_c^{open}(P', p+\frac{b}{2}+\hat m -\hat m',b+2 v_3 + \hat{s}_3'
-\hat{s}_3 )\right) 
\nonumber \\
& & 
+ \left. \left. \ Ch_c^{open}(P'=\hat P+\hat P'-\frac{i}{2}, p+\frac{b}{2}+\hat m -\hat m',b+2 v_3 +
\hat{s}_3' -\hat{s}_3) \right] \right]
\eeq
The density 
functions $\rho_1$ and $\rho_2$ are defined in equation (\ref{rhos}).
In the self-overlap of continuous or identity branes the
dependence on the label $\hat s_i$, $\hat n$ and $\hat{m}$ will drop
out.
A consequence is
that the spectrum on the brane will be independent of its
(discretized) orientation.

\subsubsection*{Set of mutually supersymmetric branes}

In order to know which set of branes preserve some supersymmetry, we have to
look at the partition function for the open strings that stretch between the
branes. The spectrum obtained must be consistent with the GSO projection. That
requires that the R-charges are odd integers\footnote{Our conventions for the
  R-charges are
  those
of \cite{Ashok:2007ui} except that we put the $U(1)_R$ charge of the minimal
model to agree with the more standard expression $Q_{MM} = \frac{s}{2} - \frac{n}{k}$.}.
Looking at the annulus amplitudes, we see that 
all the branes we defined are BPS. Moreover,
 two branes are mutually supersymmetric if and only if:
\begin{itemize}
\item The periodicity conditions are the same in each factor:
 $\hat s_0 - \hat s_i = \hat s_0' -
  \hat s_i' \ (mod\ 2)$ for $i=1,2,3$.
\item The branes satisfy the condition that angles and fluxes in various 
two-planes sum
 up to zero (extending the analysis of \cite{Berkooz:1996km}):
 \be \label{parallel} \sum_{i=0}^3 \frac{\hat s_i - \hat s_i'}{2} + \frac{(2\hat m -
   \hat n) -(2\hat 
 m' - \hat n')}{k}  = 0 \ (mod\ 2).\ee
\end{itemize}

\subsection{Semi-classical picture}

We will now discuss the semi-classical interpretation of these branes in the
 geometry (\ref{geoCFT}). It is obtained from the semi-classical analysis of
 the branes in Liouville theory (section \ref{semiclassicsLiouville}) and in
 the minimal model (section \ref{MMflat}). The only subtlety is the
 implementation of the $\mathbb{Z}_k$ orbifold in the angular direction $\omega$ 
 (which is the geometrical manifestation of the GSO projection). The
 invariance under the action
 of the $\mathbb{Z}_k$ group
 of the (regular) branes we
 consider is guaranteed by taking $k$ copies of the semi-classical branes in
 the covering space. 
In the conformal field theory we performed this step by discarding the
 Ishibashi states that do not pass the GSO projection.

The identity branes $| \hat s_0 ; \hat s_1 ; \hat j, \hat n, \hat s_2
; \hat m, \hat s_3 \rangle_{\mathbb{I}}$ are compact and located in
the more strongly coupled region.

The continuous branes $| \hat s_0 ; \hat s_1 ; \hat j, \hat n, \hat s_2
; \hat J, \hat m, \hat s_3 \rangle_c$ and the discrete branes $| \hat s_0 ; \hat
s_1 ; \hat j, \hat n, \hat s_2 ; \hat J, \hat r, \hat s_3 \rangle_d$ are
extended both in
the trumpet and in the bell, according to 
equations (\ref{extBraneLSC}) and (\ref{braneMM}). Their semi-classical
description is given by: 
\be\label{geoContBraneCFT}
\left\{ \bigcup_{n=0}^{k-1}  
\left( \cosh(r) \sin \left(
  \psi+\frac{\omega}{k}-\psi_0+\frac{2\pi n}{k} \right) = c
\ ;\ 
\sin(\theta) \sin \left( \frac{\omega}{k}-\omega_0+\frac{2\pi n}{k}
\right) =\sin{\theta_0} \right) 
\right\}
\ee
The parameters $\theta_0$ and $\omega_0$ are given in equation (\ref{paramMM}) in terms
 of the boundary state labels $\hat j$ and $\hat n$. The parameters $c$ and
 $\psi_0$ are given in terms 
 of the continuous brane labels $\hat J$ and $\hat m$ in
 equations (\ref{c(P)}) and 
 (\ref{psi0(M)}) (and for the discrete brane in equations (\ref{c(J)d}) and
 (\ref{psi0(r)d})).


\section{D-branes meet NS5-branes}\label{D4NS5}

In the previous section we built D-branes in the
coset
 background (\ref{geoCFT}). We
 now go back to the T-dual background (\ref{tdualNS5geom}) describing
 the near-horizon limit of NS5-branes on a circle. We
first interpret the
 continuous brane as a stack of D4-branes blown up by the di-electric
 effect. Using the results of
 section \ref{cft}, we
then give an exact description of the cutting of
 D4-branes into pieces as they hit NS5-branes.

\subsection{D-branes in the NS5-branes geometry}

In order to develop some intuition about the configurations we are
discussing, we first study the semi-classical description of the
branes in the
coset geometry (\ref{tdualNS5geom}).

\subsubsection*{Continuous and discrete branes}

The semi-classical description of the continuous branes is obtained from
 equation (\ref{geoContBraneCFT}). The T-duality is performed in the angular $\omega$
 direction. To get the equation defining the brane in the NS5-brane background
 (\ref{tdualNS5geom}), we proceed as in \cite{Israel:2005fn}. First we write 
the angular coordinate $\omega$ in
 terms of the other coordinates. This gives the gauge field
living on
 the brane. Then we eliminate $\omega$ from the equations
 (\ref{geoContBraneCFT}) to get the geometry of the
brane.
In a parametric form, we obtain:
\be\label{geoD4} \left\{ \begin{array}{l}
\cosh(r)\sin(\psi-\psi_0+\omega_0+\tilde{\omega})=c\\
\sin(\theta)\sin(\tilde{\omega})=\sin(\theta_0)
\end{array} \right.
\ee
\begin{figure}
\centering
\includegraphics[scale=0.82,bb=22 0 595 545]{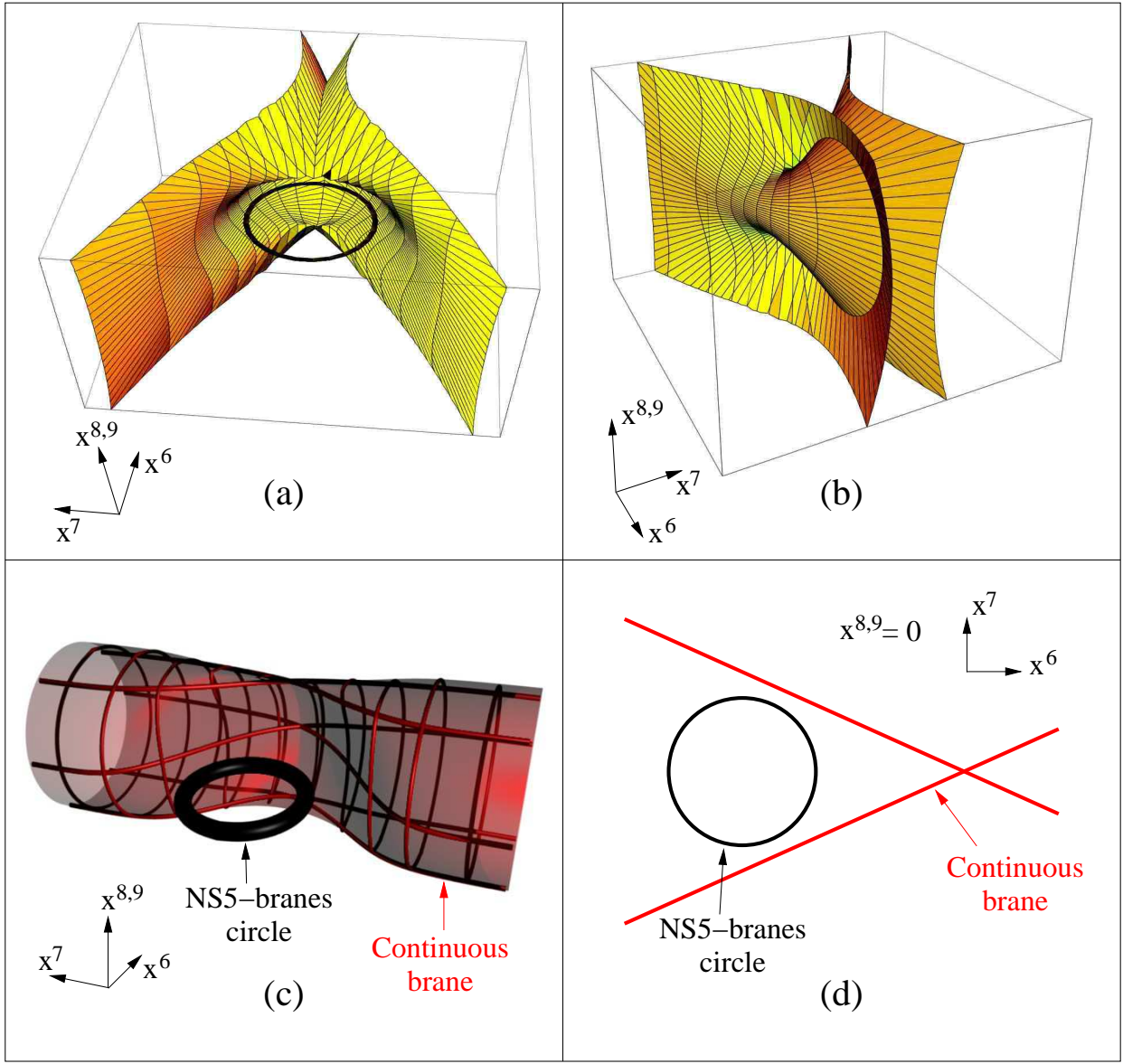}
\caption{Semi-classical picture of the single-sheeted continuous brane. 
  It is invariant under rotation in the plane $(x^8,x^9)$. The values of the
  parameters used to draw the brane are: $c=1.2$, $\theta_0=\frac{\pi}{8}$,
  $\psi_0-\omega_0=0$. The circle of
  NS5-branes is also shown. (a),(b): The
  single-sheeted continuous brane as seen from two different 
  points of view (in Euclidean space). (c) A sketch of the
  single-sheeted continuous brane, taking into account the non-trivial metric
  in space-time. (d) The intersection of the 
  single-sheeted continuous brane with the plane containing the circle of
  NS5-branes. \label{braneUncut}} 
\end{figure}
\begin{figure}
\centering
\includegraphics[scale=0.82,bb=5 0 585 545]{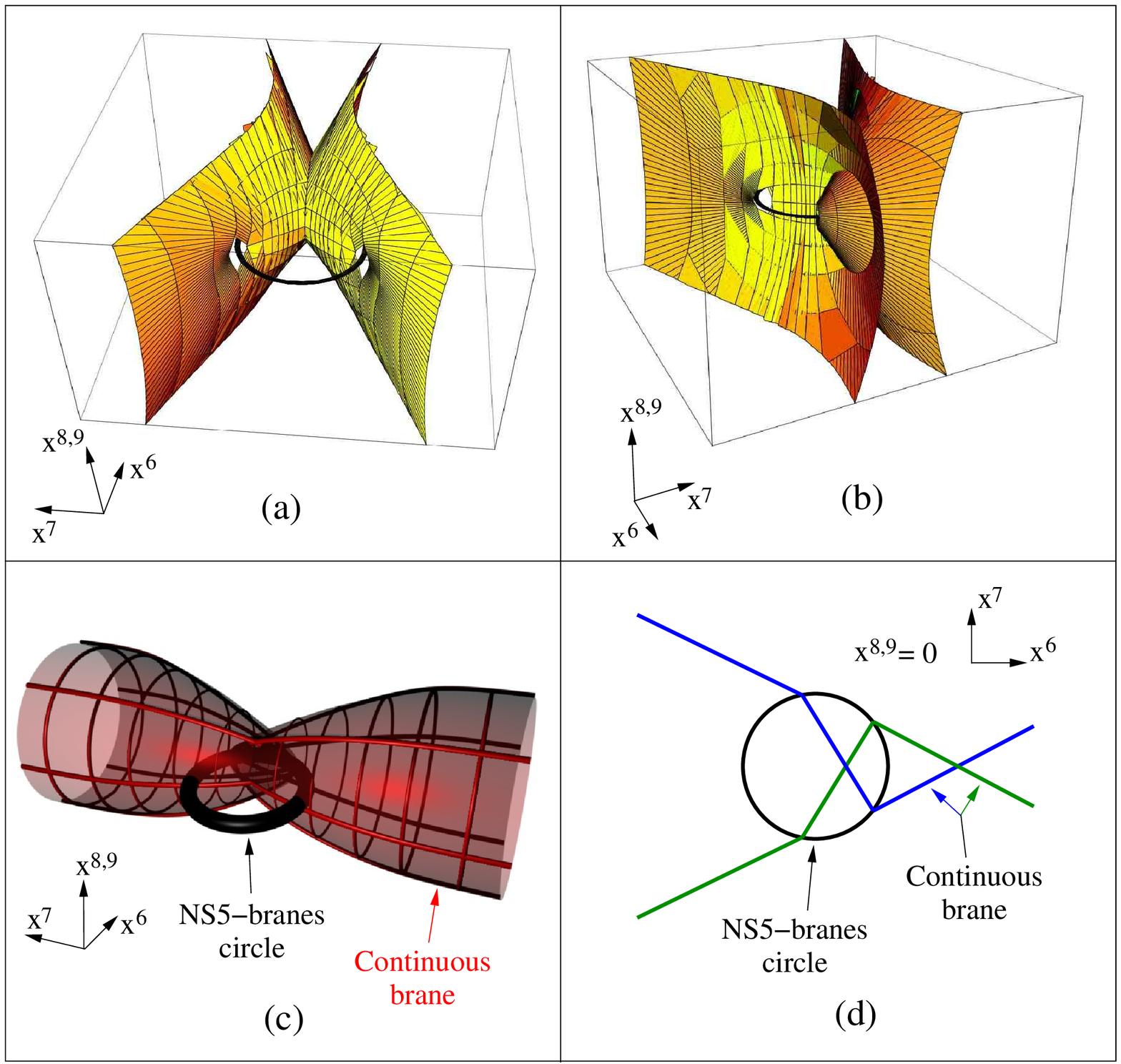}
\caption{Semi-classical picture of the double-sheeted continuous brane.
It is
  invariant under rotation in the plane $(x^8,x^9)$. The values of the
  parameters used to draw the brane are: $c=0.8$, $\theta_0=\frac{\pi}{8}$,
  $\psi_0-\omega_0=0$. The circle of 
  NS5-branes is also shown. (a), (b): The double-sheeted continuous brane as seen from 
  two different 
  points of view (in Euclidean space). (c): A sketch of the double-sheeted continuous brane,
  taking into account the non-trivial metric in space-time. (d) The intersection of the
  double-sheeted continuous brane with the plane containing the circle of
  NS5-branes. The two sheets are drawn with different colors. \label{braneCut}}
\end{figure}
The parameter $\tilde{\omega}$
 extends from zero to $2\pi$. 
The direction of the brane in the plane $(x^6,x^7)$ is given by the angle
$\psi_0-\omega_0$, which is fixed by the
combination of the brane parameters 
$\hat n - 2 \hat m$. The same combination appears in the condition of mutual
supersymmetry for two branes (equation (\ref{parallel})). This means that
mutually supersymmetric branes are parallel \footnote{If the fermionic
parameters $\hat s_i$ have different parity for the two branes, the condition
becomes that they must be orthogonal.}. The orthogonal combination $\hat n + 2 \hat
m$ gives the value of a Wilson line on the brane.
The shape of the brane is shown in figures \ref{braneUncut} and \ref{braneCut}. 
We observe that there are two physically different configurations for this
brane, according to 
the value of the parameter $c$. 
For $c > 1$, the brane does not intersect the circle of NS5-branes. 
It is
composed of a single self-intersecting sheet.
For $c<1$, the brane does intersect the circle of NS5-branes. It is
 composed of two distinct sheets. If the parameters $c$ and $\theta_0$ satisfy
 $1>c >\sin \theta_0$, the two sheets 
 intersect each other, as shown in figure \ref{braneCut}. On the other hand,
 if the parameters are such that $c< \sin \theta_0$, the two sheets do not
 intersect each other.

Notice that the continuous brane in the
 conformal field theory has positive 
parameters $\hat J$ and $\hat j$, which implies that
 the geometric parameters $c$ and $\theta_0$ satisfy: 
$c \ge \cos
 \frac{2\pi}{2k} \ge \sin \theta_0$. The conformal field theory description of
 continuous branes with parameter $c$ smaller than $\cos \frac{2\pi}{2k}$ is an
 interesting open problem.
We will make this analysis more precise in section \ref{c=1}.

For the discrete branes, the parameter $c$ is always smaller than one and 
quantized (cf. equation (\ref{c(J)d})). The semi-classical description is
otherwise 
identical.

\subsubsection*{D-branes in the presence of fluxes and the di-electric effect}

\begin{figure}
\centering
\includegraphics[scale=0.85,bb=40 30 590 290,clip=true]{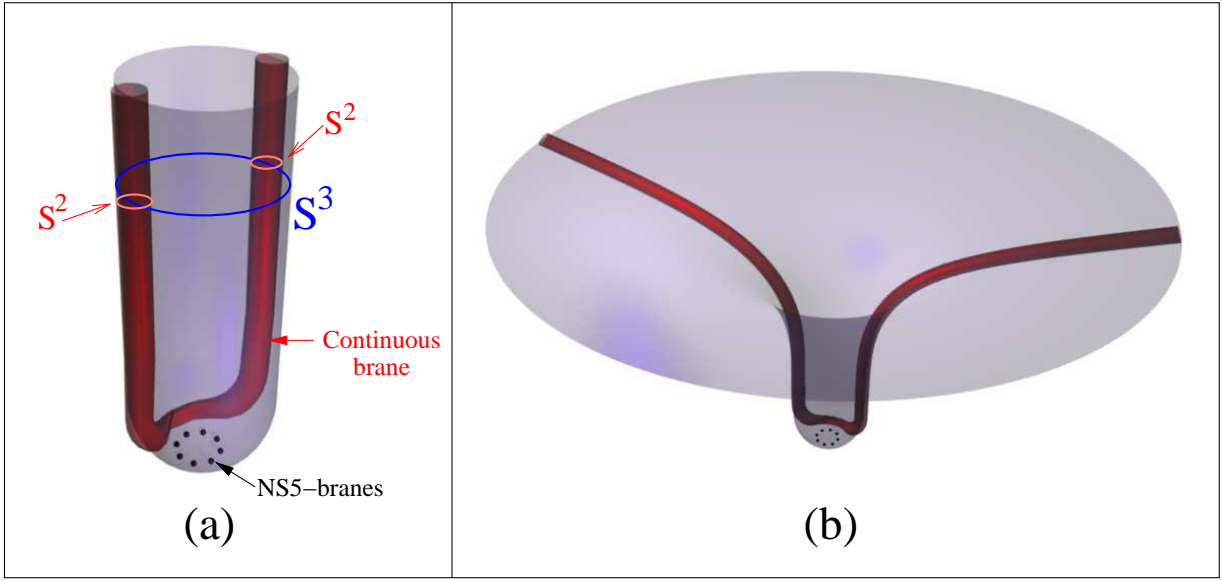}
\caption{The shape of the continuous brane far from the NS5-branes. (a) In the
  doubly scaled
  geometry (\ref{tdualNS5geom}), the intersection of the continuous brane with
  the transverse three-sphere is given by two anti-podal two-spheres. (b)
  In the full NS5-brane background (\ref{NS5background}), the continuous brane has
  two legs getting out of the throat. \label{braneThroat}}
\end{figure}

We will now argue that the continuous brane can be thought of as a stack of
D4-branes blown-up by the di-electric effect. We first discuss this effect in
the asymptotic region, 
where the doubly scaled geometry (\ref{tdualNS5geom})
 simplifies to 
 $\mathbb{R}_{(r)} \times S^3_{(\theta,\phi,\psi)}$. The radius of the
 three-sphere is equal to $\sqrt{k \alpha'}$. This is the geometry of
 the throat
 created by $k$ 
NS5-branes. In this region the geometry of the 
continuous
 brane (\ref{geoD4})
 simplifies to:
\be\label{asympGeoD4} \left\{ \begin{array}{l}
r \gg 1 \\
\sin(\theta) \sin(\psi-\psi_0) = \pm \sin(\theta_0)
\end{array} \right. \ee
The second line defines two anti-podal two-spheres embedded in the three-sphere
 $(\theta,\phi,\psi)$. The radius of these two-spheres is equal to $\sqrt{k
 \alpha'} \cos \theta_0$.
We deduce that the geometry of the continuous brane far from the circle is
$(S^2+S^2)\times \mathbb{R}$, as drawn in figure \ref{braneThroat} (a).
So in the full NS5-brane background
 (\ref{NS5background}), the tube goes into
and out
 of the throat  (cf. figure \ref{braneThroat} (b)).

D-branes of low dimensional world-volume in the presence of fluxes tend to
develop additional world-volume dimensions (see \cite{Myers:1999ps} for the
case of RR fluxes and \cite{Alekseev:2000fd} for the case of NSNS fluxes). This
is a di-electric effect. In particular in \cite{Alekseev:2000fd} it
was shown that 
$n$ coincident D0-branes in the $SU(2)_k$ WZWN model tend to decay to 
a single spherical D2-brane. The radius of this brane is $\sqrt{k \alpha'}
\sin(\pi n/k)$ \cite{Maldacena:2001ky}.  

The same phenomenon arises in the NS5-brane background (\ref{tdualNS5geom}), since the NS5-branes
source NSNS-flux 
on the three-sphere
$S^3_{(\theta,\phi,\psi)}$. So even 
though the world-volume of the continuous
brane (\ref{geoD4}) 
is seven-dimensional (three-dimensional in the
transverse space 
$x^{6,7,8,9}$, and also extended in four flat directions), we can also think of this single extended D6-brane as
a stack of $n$ 
coincident D4-branes. We identify the number $n$ of D4-branes via the radius
of the two-spheres defined by equation (\ref{asympGeoD4}):
\be n = 2 \hat j +1 \ee

The D-branes with one dimension in the transverse space $x^{6,7,8,9}$ and extremizing
the DBI action in the NS5 background (\ref{NS5background}) are
straight lines in the $x^{6,7,8,9}$ coordinates.
 When we minimize the
tension as well as the length, the harmonic function drops out in the calculation. Since
the brane (\ref{geoD4}) is invariant under rotation in the $(x^8,x^9)$
plane, we consider only D-branes that preserve that symmetry. They are
straight lines in the $(x^6,x^7)$ plane,
located at the origin of the $(x^8,x^9)$ plane. The direction of this
line is given by the angular parameter $\psi_0-\omega_0$, and the coordinate distance to the
origin is given by the parameter $c$ multiplied by $\rho_0$.

 In fact,
we are considering the bulk and brane system in a 
{\em triple } scaling
limit, which consists of the usual double scaling limit in the bulk,
and a third limit that scales the coordinate distance of the branes to
the NS5-branes to zero, commensurate with the scaling down of the
coordinate distance between the NS5-branes, such that the ratio (equal
to the parameter $c$) is fixed.

\begin{figure}
\centering
\includegraphics[scale=0.6]{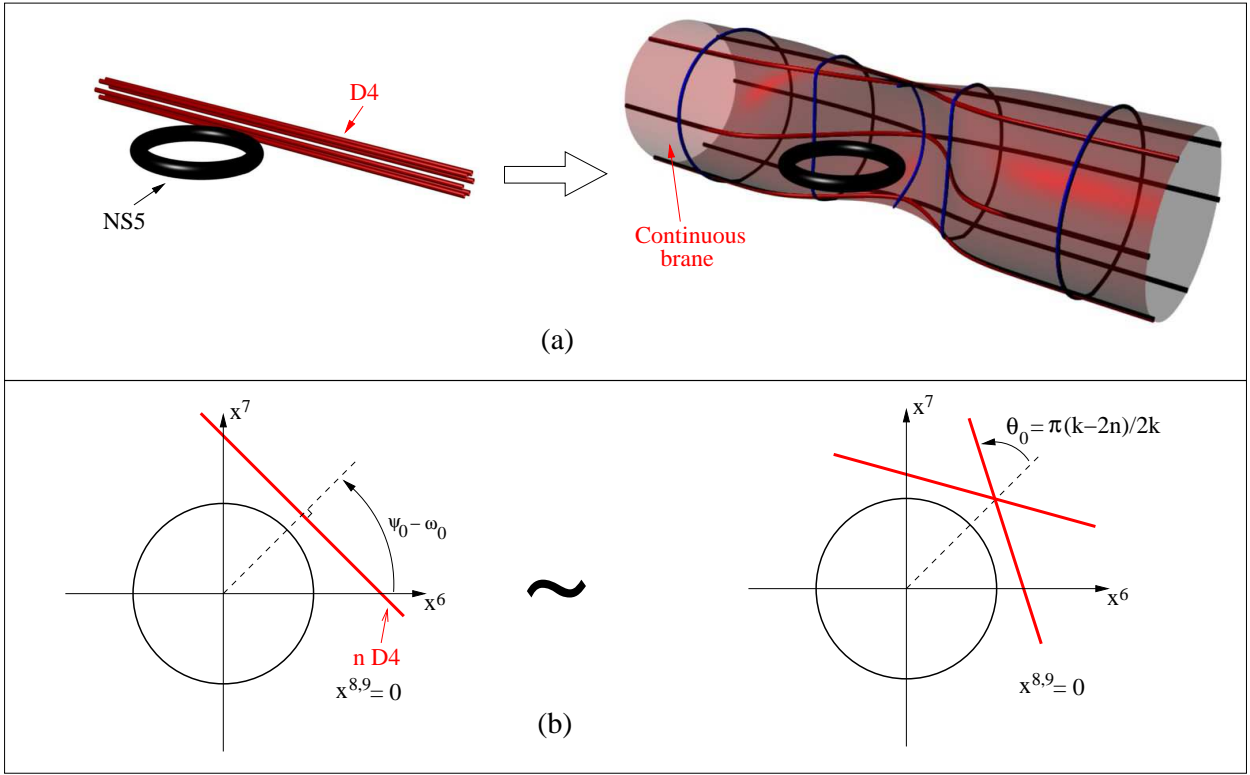} 
\caption{The di-electric effect for a stack of D4-branes in the
  presence of a circle of NS5-branes. (a) The coincident D4-branes are
  blown up to a single continuous brane. Far from the circle, the continuous brane has
  geometry $(S^2+S^2)\times \mathbb{R}$ in the transverse space
  $x^{6,7,8,9}$. Close to the circle, the two-spheres are deformed by
  the repulsion between the D6-brane and the NS5-branes. (b) The
  intersection of the previous D-branes with the plane containing the
  NS5-branes. We indicate how the parameters defining the continuous brane and
  the straight D4-branes are related. \label{myers}}
\end{figure}

We now understand that equations (\ref{geoD4}) describe the
manifestation of the di-electric effect for D4-branes in the presence
of a circle of NS5-branes. A stack of coincident D4-branes decays to a
single D6-brane which shape is given by (\ref{geoD4}) (cf. figure
\ref{myers}).  Far from the circle, the D6-brane has geometry
$(S^2+S^2) \times \mathbb{R}$ in the transverse space
$x^{6,7,8,9}$. Close to the NS5-branes, the two-spheres are deformed
by the repulsion between the NS5-branes and the D6-brane. We can
understand the way the two-spheres are deformed as follow. Consider
the two-sphere on one leg of the brane. The north pole of the sphere
is 
pulled inwards as we get closer to the NS5-branes. It reaches the
south pole and goes beyond as the distance to the NS5-branes increases
again. We end up on the second leg with the second two-sphere. The
two-sphere changes orientation in the process.

In more detail, consider for example the case where the D4-branes do
not intersect the circle of NS5-branes ($c>1$). The $\mathbb{Z}_2$
symmetry $\psi \to -\psi+2\psi_0-2\omega_0$ has a fixed point at
$\psi=\psi_0-\omega_0$: this is the point where the D4-branes are
closest to the circle of NS5-branes. This particular angular direction
also defines the locus of the self-intersection of the continuous
brane (\ref{geoD4}). In terms of the coordinates $x^{6,7,8,9}$ defined
in (\ref{cartcoords}), this surface has equation:
\be\label{intersectD3} \left\{ \begin{array}{l} -\sin \psi_0 x^6 +
\cos \psi_0 x^7=0\\ \left( \cos \psi_0 x^6 + \sin \psi_0 x^7
-\frac{\rho_0}{2}\left( \frac{c}{\sin \theta_0} + \frac{\sin
\theta_0}{c} \right)\right)^2 + (x^8)^2 + (x^9)^2 = \rho_0^2 \left(
\frac{1}{4} \left( \frac{c}{\sin \theta_0} + \frac{\sin \theta_0}{c}
\right)^2 -1 \right) \\ \cos \psi_0 x^6 + \sin \psi_0 x^7 \ge c \sin
\theta_0
\end{array} \right. \ee
The first two lines describe a surface with topology of a two-sphere,
while the third line removes a piece of this surface. Actually, the
surface is double-sheeted and without boundary. Indeed, this surface
describes the deformed two-sphere of the continuous brane at the point
where the north pole meets the south pole and the sphere changes
orientation (cf. figure \ref{myers}).

When the stack of D4-branes intersects the circle, with
parameters such that $1 \ge c \ge \sin \theta_0$, the D6-brane has two sheets that
intersect each other along the surface defined by the same equations
(\ref{intersectD3}). However this time the inequality is always satisfied, and
we have
a full two-sphere. If the parameters satisfy
$c < \sin \theta_0$, the two
sheets of the brane do not intersect each other anymore.

\subsubsection*{Identity branes}
To obtain a semi-classical description of the identity branes, it is
convenient to consider them in the background mirror
 to
 (\ref{geoCFT}), where the trumpet is replaced by the cigar and the region
near the tip  is under geometric control. Building the identity branes in the
conformal
field theory mirror to
 (\ref{fullCFT}), we can deduce the semi-classical description of the identity
 branes in the Neveu-Schwarz five-brane background
 (\ref{tdualNS5geom}) \cite{Israel:2005fn}. 
We can also use the following semi-classical description of the identity brane
in the trumpet geometry (\ref{trumpet}):
\be r=0 \ ; \ \psi-\psi_{\mathbb{I}}=0 \ \ee
with the angular direction $\psi_{\mathbb{I}}$ defined in terms of the brane parameter
$\hat m$ as:
\be \psi_{\mathbb{I}} = 2\pi \frac{\hat m}{k}.\ee 
Then we perform a T-duality as in the previous
case to get the
semi-classical description of the identity branes in the NS5-brane background:
\be \label{geoD4suspended} 
\left\{ \begin{array}{l} 
r=0 \\
\sin(\theta) \sin \left( \psi - \psi_{\mathbb{I}} + \omega_0 \right) =\sin(\theta_0)
\end{array}
\right.
\ee
The parameters $\theta_0$ and $\omega_0$ are defined in terms of the brane
parameter $\hat j$ and $\hat n$ in (\ref{paramMM}).
The parameter $\hat j$  fixes the length of the brane.
These branes are straight lines in the $(x^6,x^7)$ plane, 
starting and ending on the circle of NS5-branes. 
As for the continuous branes, the combination of the brane parameters $\hat n - 2 \hat m$
gives the 
angular orientation of the brane.

\begin{figure}
\centering
\includegraphics[scale=0.8]{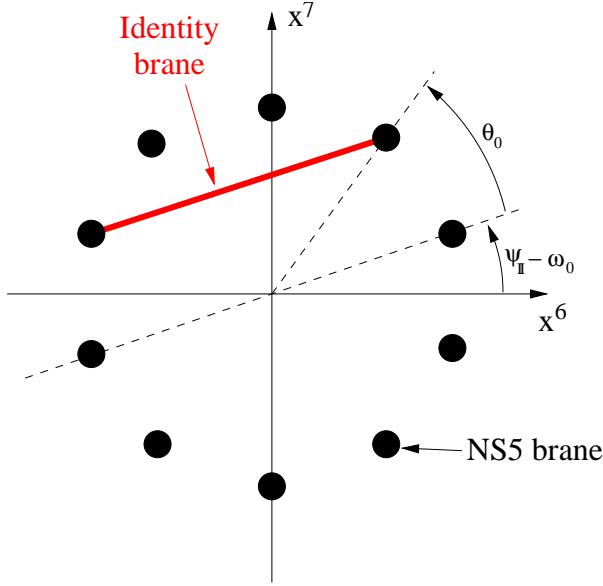}
\caption{Semi-classical description of an identity brane, stretching between
  NS5-branes.\label{braneId}}
\end{figure}

In the minimal model, the choice of parameters $\hat j$, $\hat n$, $\hat s_2$
is constrained by the selection rule  (namely:
$2 \hat j + \hat n + \hat s_2$ is zero modulo two)\footnote{For simplicity we
  only discuss the
branes of the minimal model with $\hat s_2$
  even. The 
branes with $\hat s_2$ odd have a
  semi-classical interpretation that differs from the one given in equation
  (\ref{braneMM})  
(see \cite{Maldacena:2001ky}).}. As a consequence, the
length of an identity brane is related to its orientation. 
Note that even though the NS5-branes have been spread on the circle, their
position is still coded in the worldsheet instantons. We make the minimal
assumption that the identity branes are suspended between NS5-branes.
Since the identity branes
are defined only for integer values of the parameter $\hat m$, this assumption
is consistent with the selection rule in the minimal model.
The semi-classical description of the identity branes gives a 
way to recover the 
positions of the NS5-branes on the circle.

\subsection{D-branes cut in two pieces}\label{c=1}
Consider a continuous brane (\ref{cBrane}).
In order to obtain a semi-classical picture of the operation we performed in the 
exact boundary conformal field theory, we study what happens when the brane parameter
 $\hat J$
 goes to one half (i.e. the imaginary part of $\hat J$ goes to zero), keeping
 all the other brane parameters fixed.
 The geometric parameter $c$ goes down to one as the brane parameter $\hat J$
 goes to one-half. 
 When the geometric parameter $c$ is equal to one, the continuous brane touches the
 circle and is 
on the verge of being cut into two 
 intersecting sheets.

Let's make this point more precise. First notice that the surface
(\ref{geoD4}) is symmetric under the reflection $\psi-\psi_0+\omega_0 \to
-(\psi-\psi_0+\omega_0)$ (together with $\tilde{\omega} \to \pi - \tilde{\omega}$). 
Figures \ref{braneUncut} and \ref{braneCut} suggest that when the brane is cut, the two sheets are
exchanged by this symmetry.
We would like to know whether there exists a smooth path $\gamma$ on the brane
connecting 
two symmetric points, 
of respective angular coordinate $\psi_1$ and $-\psi_1+2\psi_0-2\omega_0$.
The value of the parameter $\tilde{\omega}$ for these points is noted
respectively 
$\tilde{\omega}_1$ and $\pi-\tilde{\omega}_1$. Along the
path $\gamma$, the angular coordinate $\psi$ goes smoothly from $\psi_1$ to
$-\psi_1+2\psi_0-2\omega_0$, and the parameter $\tilde{\omega}$ goes smoothly from
$\tilde{\omega}_1$ to $\pi-\tilde{\omega}_1$. Notice that the parameter
$\tilde{\omega}$ has to stay within the range  $\theta_0 \le \tilde{\omega} \le
\pi-
\theta_0$ for the equations (\ref{geoD4}) to have a solution. For the same
reason the sum 
$\psi-\psi_0+\omega_0+\tilde{\omega}$ must be positive and smaller than $\pi$. We
deduce that along the path $\gamma$, the sum 
$\psi-\psi_0+\omega_0+\tilde{\omega}$ goes smoothly from
$\psi_1-\psi_0+\omega_0+\tilde{\omega}_1$ to $\pi-(\psi_1-\psi_0+\omega_0+\tilde{\omega}_1)$,
necessarily crossing the value $\psi-\psi_0+\omega_0+\tilde{\omega}=\frac{\pi}{2}$. But
this is possible if and only
if 
the brane parameter $c$ is greater or equal to
one. Indeed, if $c<1$,
the first equation in (\ref{geoD4}) does not have solutions for
$\psi-\psi_0+\omega_0+\tilde{\omega}=\frac{\pi}{2}$.
The conclusion is that the
path $\gamma$ exists for $c \ge 1$. In that case, the two points are
smoothly connected and the brane 
is single-sheeted. If $c<1$, there is no smooth path between two symmetric
points, and the brane is made of two 
distinct sheets.
 If $1>c>\sin \theta_0$, the two sheets of the brane
intersect and one can find a path between two symmetric points. However the
parameter $\tilde{\omega}$ will not vary continuously along such a path,
 indicating
that we pass from one sheet onto the other.

\begin{figure}
\centering
\includegraphics[scale=0.85]{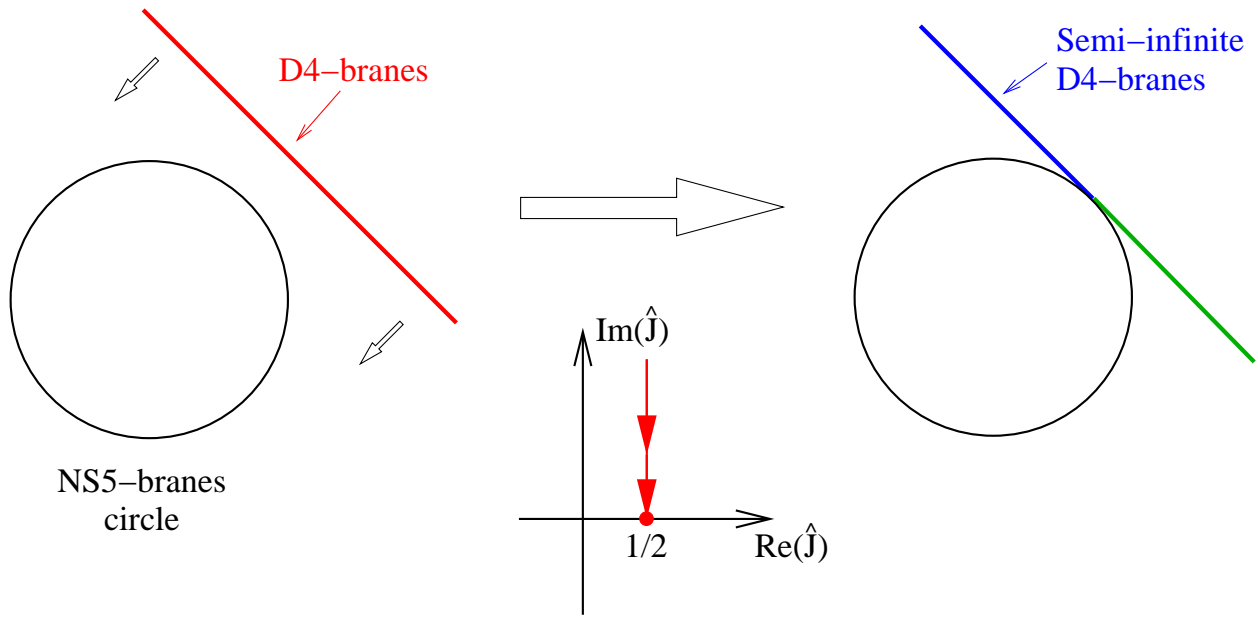}
\caption{A stack of D4-branes approaches a circle of NS5-branes, as the brane
  parameter $\hat J$ goes to one-half. For $\hat J=\frac{1}{2}$ the D4-branes are cut in 
  two pieces.\label{cutIn2}}
\end{figure}

We argued in the previous section that the continuous
D6-brane can be interpreted as a stack of D4-branes.
The cutting of the D6-brane in two sheets can also be understood in terms of
the D4-branes. As the D4-branes hit the NS5-circle, they are cut in two
semi-infinite pieces 
(see figure \ref{cutIn2}).

The brane addition relation (\ref{c=d+d}) is valid precisely at the point where the
D6-brane is cut in two pieces.
 That implies that the continuous brane with
parameter $\hat J = \frac{1}{2}$ (and $\hat m$ half-integer) is the sum of two discrete branes.
The picture obtained in the conformal field theory fits nicely with the
semi-classical analysis.

\subsection{D-branes cut in three pieces}
Now we want to study the continuous brane when it gets inside
 the circle of NS5-branes, that is for values of the
 parameter $c$ smaller than one.
Consider a continuous brane (\ref{cBrane}) with parameter
$\hat m$ integer (and any value for the other
 parameters).
We study what happens when the parameter
 $\hat J$
 goes to zero, keeping all the other parameters fixed. 
That mirrors  what we did in section \ref{bound}, while computing the
overlap of continuous branes in Liouville
 theory.
The addition  relation (\ref{c=i+d+d}) implies that an identity
 brane (\ref{IBrane}) appears when the continuous brane parameter
 $\hat J$ is equal to zero. The
 identity brane created has the same parameters as the continuous brane.
We will now understand the appearance of this identity brane from the semi-classical
analysis of the continuous branes. 

\begin{figure}
\centering
\includegraphics[scale=0.75]{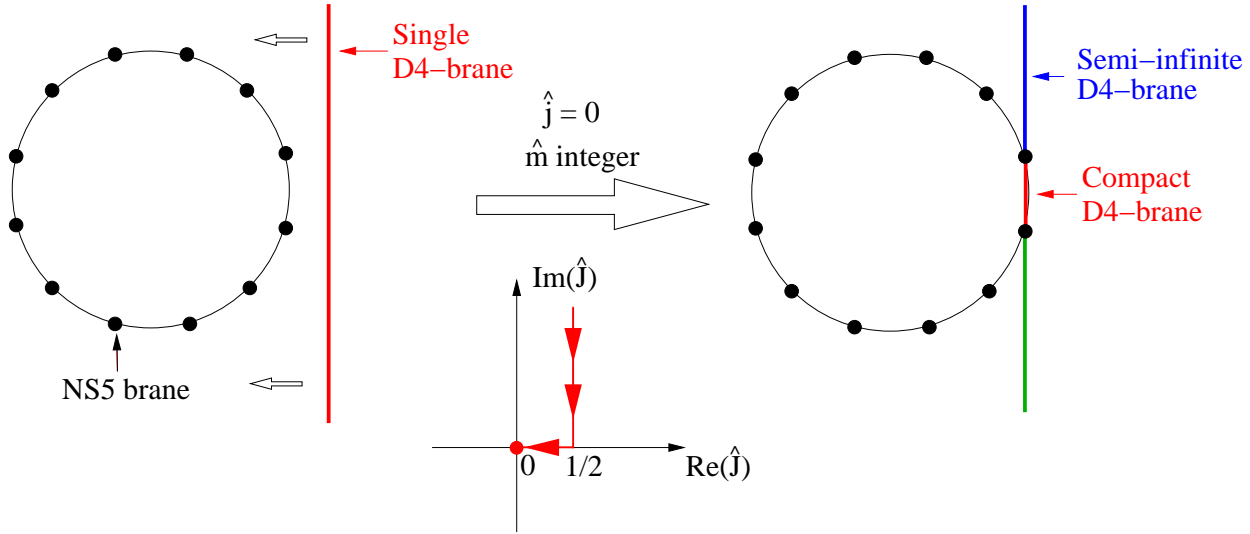}
\caption{A particular case: a single D4-brane approaches a circle of
  NS5-branes, as the brane 
  parameter $\hat J$ goes to zero (the parameter $\hat m$ is integer). For $\hat J=0$
  the D4-brane is cut in 
  three pieces. Massless open strings live on the compact piece.\label{cutIn3case1}}
\end{figure}

Let's first consider the simpler case
in which the continuous brane
parameter $\hat j$ is zero. In this situation, the continuous D6-brane represents
one single D4-brane.
 Now we observe that at the
special value of the brane parameter $\hat J=0$, the geometric parameter $c$
is equal to $\cos(\frac{2\pi}{2k})$: we are considering the situation where the
D4-brane hits a pair of neighboring NS5-branes (cf. figure
\ref{cutIn3case1}). The compact part of the D4-brane that stretches between the
NS5-branes matches precisely the semi-classical description
(\ref{geoD4suspended}) of the
identity brane on the right-hand side of equation (\ref{c=i+d+d}). 
We conclude that
 in this case the brane addition relation (\ref{c=i+d+d})
 captures the 
fact that the D-brane
hitting two separated NS5-branes consists of three pieces.

\begin{figure}
\centering
\includegraphics[scale=0.75]{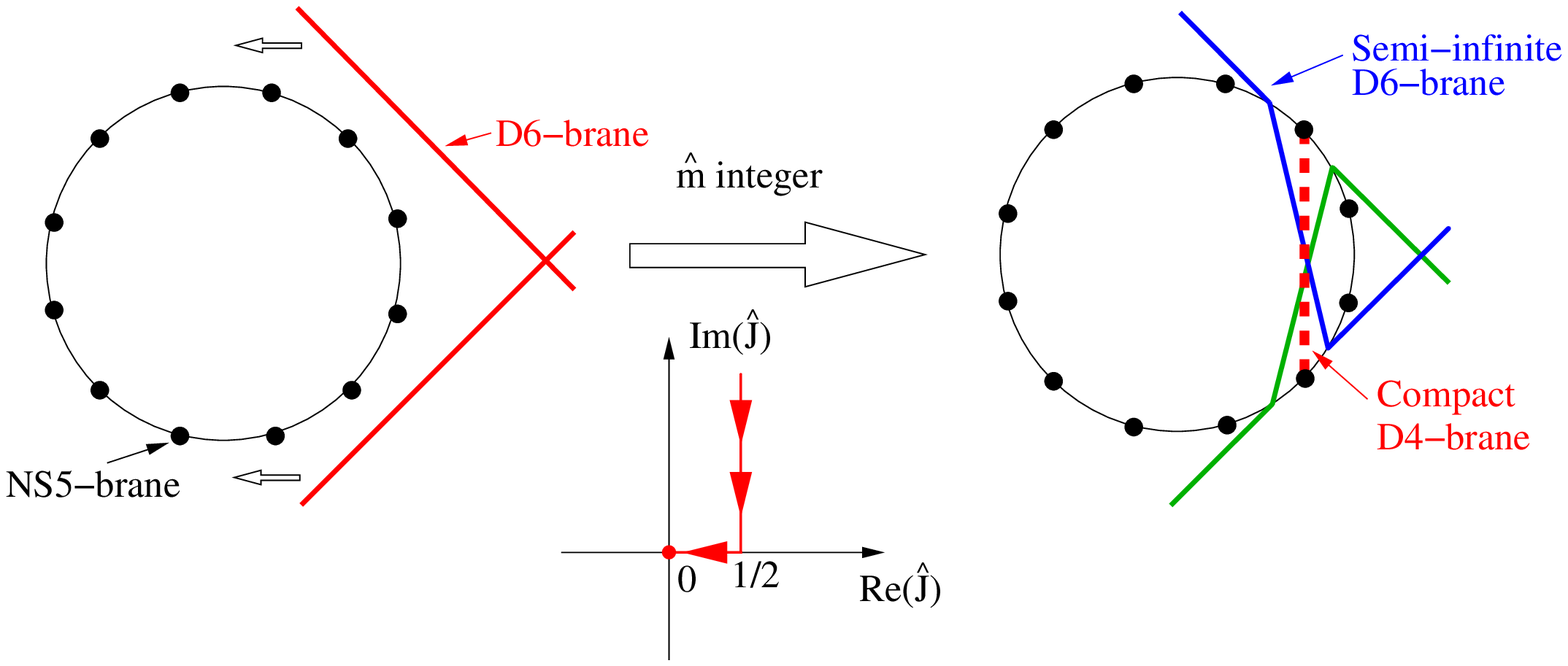}
\caption{The generic case: a continuous brane approaches a circle of
  NS5-branes, as the brane 
  parameter $\hat J$ goes to zero (the parameter $\hat m$ is integer). For $\hat J=0$
  an identity brane appears, sketched by a dotted line. Massless open strings
  live on this compact 
  brane.\label{cutIn3case2}} 
\end{figure}

Next, let's consider the generic case where the brane parameter $\hat j$ is
non-zero. The identity brane that appears 
does not stretch between
 neighboring
NS5-branes. To understand the appearance of such a compact brane, we have to
take into account the 
blowing 
up of the stack of D4-branes by the 
di-electric
effect. In figure \ref{cutIn3case2} we draw the intersection of the continuous
brane (\ref{geoD4}) with the plane containing the circle of NS5-branes. We observe that at
the special value of the brane parameter $\hat J=0$, the intersection of the
continuous brane with the interior of the circle of NS5-branes matches 
(up to corrections of order $\frac{1}{k}$) the semi-classical description (\ref{geoD4suspended}) of the identity brane 
that is generated. We are still describing 
D-branes that are cut into three pieces when
hitting two separated NS5-branes. But the way the
D-branes are cut is less intuitive,
 because they are blown up by the di-electric effect.

The computation of section
 \ref{bound} tells us that when the parameter $\hat J$ is equal to
 $\frac{1}{4}$, new massive states appear
 on the continuous brane. They are bound states on the world-volume on the
 brane. The mass of these states goes down as the parameter $\hat J$ goes to
 zero.
At the special value of the parameter $\hat J=0$, some of these localized
 states become massless.  
  The new massless modes live on the identity
 brane of the addition  relation (\ref{c=i+d+d}). So the appearance of poles
 in the densities of states studied in section \ref{bound} corresponds to the
gradual 
creation of a new brane in space-time.

Another way to see this is to look at the $D0$-charge on the
 continuous $D2$-brane in the cigar. The $D0$-charge is defined as:
 $\frac{2}{2\pi}\int F=\frac{2k}{2\pi}\arcsin c $. The additional
 factor of two comes from the fact that the brane is
 double-sheeted. As the parameter $c$ goes from 1 to $\cos
 \frac{2\pi}{2k}$, the $D0$-charge changes by one unit. In our
 context, this corresponds to the creation of one unit of D4-brane
 charge.

The fact that the addition relation (\ref{c=i+d+d}) is only valid for integer values of
the brane parameter $\hat m$ is consistent with the position of the NS5-branes
on the circle, and thus with the selection rule in the minimal model.

We repeat that the movement of the continuous brane as the parameter $\hat J$ changes is
infinitesimal when measured in the asymptotic flat 
space metric
 (due to the triple scaling). Indeed the geometry of the continuous brane far from the circle
(given by equations (\ref{asympGeoD4}))
does not depend on the parameter $\hat J$. 

It would be interesting to study what happens when the continuous brane gets even
closer to the center of the NS5-branes circle. However making the analytic
continuation of the continuous brane parameter $\hat J$ to negative values is
not the right way to proceed, since it induces the appearance of non-unitary
characters of the $N=2$ superconformal algebra. An alternative could be to
introduce a non-factorized brane as the continuation of the continuous brane
beyond $\hat J=0$.


\section{Four-dimensional gauge theory in the depths of the throat}\label{gauge}

The discussion of the previous section is valid independent of the
 boundary conditions we
choose for the flat space directions (provided we have 
 at least two directions with
Dirichlet boundary conditions). In this section we focus on D4-branes,
 extended in the $x^{0,1,2,3,6}$ directions  and localized in the 
other directions, as well as  mutually supersymmetric
D6-branes. We perform an analysis of 
 the low-energy excitations on these branes, and a preliminary analysis of
 their
interactions.

\subsection{Energy scales}
First we 
specify
the energy scales on which we will focus.  There are two
fixed energy scales in the problem, namely the string scale $M_s =
1/\sqrt{\alpha'}$ and the curvature scale $M_{gap} = 1/ \sqrt{k \alpha'}$. At
the string scale  oscillation modes become important. The curvature
scale gives the order of the mass gap due to the linear dilaton, as well as
the Kaluza-Klein mass gap due to compactification of some of the transverse
directions.  The low energy scales we focus on are energies far below both the
string scale and the curvature scale.

\subsection{Localized gauge theory on the non-compact brane}

Let's consider the theory living on a continuous brane of parameter $\hat{J}$.
In the throat, due to the presence of the linear dilaton, the theory on a
generic continuous brane is gapped, and at energies below $M_{gap}$ we have a
theory with no excitations at all.  Operationally, if $\hat{J}$ has an imaginary
part, or we have real $\hat{J}$ in the range $1/4<\hat{J} \le 1/2 $ , the self-overlap of
the continuous brane is given in
equation
(\ref{Z_cc0}) and all states are in massive representations. The theory at low
energy is trivial in this case.

When we bring down the brane parameter into the range $0<\hat{J} < 1/4$, the
self-overlap of the continuous brane is given in equation (\ref{Z_cca}). All
the states are still in massive representations, but we notice that a state
localizes and its mass drops below the mass gap $M_{gap}$. By tuning $\hat{J}$
to be close to $\hat{J}=0$, we can make the mass parametrically small compared
to $M_{gap}$.  At energies below that mass, we still have a trivial theory. We
may get an interesting theory at energies squared between 
$ 2 \hat{J} (1-2\hat{J}) M_{gap}^2$ and $M_{gap}^2/4$.

At these energy scales, the theory is localized in the transverse space and is
effectively four-dimensional. It has eight supercharges.
Via the annulus amplitude we identify the excitations of the theory and they
fill out  a single massive gauge multiplet of
four-dimensional ${\cal N}=2$ supersymmetry.
We find a real massive vector, four Majorana spinors 
and five scalars. If we denote the normalizable bound state in the full
theory 
(at $j=0=n$) by:
\begin{eqnarray}
|k_{\mu=0,1,2,3},0,[\hat{J}] \rangle
\end{eqnarray}
then the excitations filling out the massive gauge multiplet are:
\begin{eqnarray}
& & \psi^{2,3}_{-1/2} |k_\mu,0,[\hat{J}] \rangle
\nonumber \\
& & \psi^{4,5}_{-1/2} |k_\mu,0,[\hat{J}] \rangle
\nonumber \\
& & G_{-1/2}^{\pm,nc} |k_\mu,0,[\hat{J}] \rangle
\nonumber \\
& & Sp_{\mp 1} G_{-1/2}^{\pm,nc} |k_i,0,[\hat{J}] \rangle.
\end{eqnarray}
We have used the superscript $nc$ to denote an operator acting on the
non-compact $N=2$ Liouville theory only and the notation $Sp_{\mp 1}$  indicates
a spectral flow by $\mp 1$ units in both the minimal model and the non-compact
model (at total central charge $c=6$).  For instance, starting from the state
$G_{-1/2}^{+,nc} |k_\mu,0,[\hat{J}] \rangle$ we can generate another state
of the same conformal dimension
by spectrally flowing by minus one unit. The state we obtain is anti-chiral in
the minimal model, 
with R-charge $-1+2/k$. It has R-charge $-2/k$ in the
non-compact model for a  
total R-charge of $-1$. 
We refer to appendix \ref{k=3} for an explicit analysis of how these states
are coded in the annulus partition function in the particular case of $k=3$
NS5-branes. 

The transverse polarizations of the massive vector are given by the first two
states, while the other states give the longitudinal polarization as well as
five scalar fields. 
We can figure out which scalar plays the role of the
longitudinal polarization of the vector by the following reasoning.
 In an old covariant approach to the determination of the spectrum, we have a
 null state in the Hilbert space which corresponds to $G_{-1/2}^{N=1}
 |k_{\mu},0,[\hat{J}] 
 \rangle$. It is a null state since $G^{N=1}$ is the 
supercurrent corresponding to the gauging of
$N=1$ supergravity on the world-sheet. 
Explicitly, we have that this null state is:
\begin{eqnarray}
k_\mu \psi^\mu_{-1/2} + G^{N=1,nc}_{-1/2} |k_\mu, 0, [\hat{J}] \rangle,
\end{eqnarray}
where we used the fact that the supercurrent annihilates the vacuum in the
Dirichlet and the minimal model factors.
Thus in the light-cone gauge where we broke Lorentz invariance and chose to 
fix the gauge symmetry by gauging away the light-cone excitations, we identify the scalar
\begin{eqnarray}
G^{N=1,nc}_{-1/2} |k_\mu,0,[\hat{J}] \rangle
\end{eqnarray}
as the longitudinal component of the massive vector field. The other five
scalars complete the massive ${\cal N} =2$ vector multiplet.

When we proceed to consider a brane with parameter $\hat{J}=0$, we find that the
massive vector multiplet splits into a massless vector multiplet and a 
massless hyper multiplet.  
By the previous reasoning, and the fact that $|k_\mu,0,[0] \rangle$ is annihilated
by both $G^{\pm,nc}_{-1/2}$, we find that the longitudinal polarization of the
vector is now null (as it must be for a massless vector). There is then only
one possible set of multiplets of the supersymmetry algebra corresponding to
the vector and the six scalars, namely a massless vector and hyper multiplet.
Explicitly, the massless vector multiplet has bosonic degrees of freedom:
\begin{eqnarray}
& & \psi^{2,3}_{-1/2} |k_\mu,0,[0] \rangle
\nonumber \\
& & \psi^{4,5}_{-1/2} |k_\mu,0,[0] \rangle
\end{eqnarray}
and the massless hyper multiplet has four bosonic scalar degrees of freedom:
\begin{eqnarray}
& &  |k_\mu,0,\pm 1/2 \rangle
\nonumber \\
& &  Sp_{\mp 1} |k_\mu,0,\pm 1/2 \rangle.
\end{eqnarray}
We indicated the discrete representation chiral and anti-chiral ground states
by $|0,\pm 1/2 \rangle$.
Alternatively, at $\hat{J}=0$ we can interpret the excitations as coming from various overlaps
of terms in the brane addition relation (\ref{c=i+d+d}). The four bosonic degrees of freedom
in  the ${\cal N}=2$ vector multiplet arise from the 
self overlap of the
identity brane on the right-hand side of the addition relation (\ref{c=i+d+d}). The other massless
states come from the overlap of the identity brane with the discrete
branes. This overlap contains the discrete characters of (\ref{1stIdEx}). 
In the character $Ch_d(J=\frac{k}{2},r=0)$ we find one chiral
primary state with $U(1)$ R-charge $Q=1$, and the spectral-flowed anti-chiral
primary with $Q=-\frac{k}{2}$. These states combine with chiral and
anti-chiral primary from the minimal model and the flat direction to give
massless modes (with $Q=\pm 1$). 
In the second discrete character $Ch_d(J=1,r=-1)$ we find one anti-chiral
primary state with $Q=-1$, and the spectral-flowed chiral
primary with $Q=\frac{k}{2}$. 
Notice that from the right-hand side of (\ref{c=i+d+d}), we expect
\textit{eight} massless bosonic states from the overlap of the identity brane with the
two discrete branes. However half of them are canceled by the self overlap of
the discrete branes, as shown in appendix \ref{discretecouplings}. 

\subsection*{Interactions}
In the spectrum of excitations, 
we witnessed a reverse Higgs mechanism consistent with ${\cal N}=2$
supersymmetry in four dimensions. 
When we survey what the relevant
Higgs mechanism can be, consistent with ${\cal N}=2$ supersymmetry in four
dimensions (see \cite{Fayet:1978ig}), we find only one candidate. A real massive vector multiplet arises
when we add a Fayet-Iliopoulos D-term to a $U(1)$ theory with a charged
hyper multiplet. The vector multiplet and the hyper multiplet acquire 
 equal
masses (proportional to the charge times the 
FI-parameter). Indeed, we did find that as we change $\hat{J}$, the vector and
hyper multiplet acquire 
the same mass. 

To nail down this picture we
need to analyze the interactions between open string states to verify
the presence of the necessary cubic and quartic interaction terms. That may be 
doable on the basis of the results of \cite{Hosomichi:2006pz} that map
boundary $n$-point functions in $H_3^+$ into an in principle
solved problem in boundary bosonic Liouville theory. One can apply this
reduction to boundary $N=2$ Liouville theory and thus determine the open string
interaction terms. We will not attempt to perform this calculation here.
It would be interesting to do it, and in particular to prove explicitly
that at non-zero $\hat{J}$ the hyper multiplet is indeed charged.

Without the necessary study of the interaction terms we can make one more
observation that is consistent with the Higgs mechanism advocated above, which
is the breaking of the (accidental) R-symmetry (at low energy). When the brane
parameter $\hat{J}$ is zero, the Hilbert space of open strings splits into
identity and discrete sectors. The scalars in the hyper multiplet are at the
lowest level available in their sector, and there are no null states at the
lowest level.
Moreover, the action of the supersymmetry generators pairs up the scalars
into doublets of the $U(2)_R$ symmetry of the theory. Each scalar pairs up
with its spectral flowed counterpart.
On the other hand, when the brane parameter is non-zero, the low-energy symmetry that exchanged
the two scalars does not commute anymore with the BRST operator (as can
clearly be seen by the existence of the null state that we mentioned
above). The $U(2)_R$ symmetry gets broken to a $U(1)_R$ that does not act on
the component of the complex doublet in the hyper that acquires a vacuum
expectation value.

Finally, we note that {\em at} $\hat{J}=0=\hat m$ we have that a marginal deformation of
type F on the world-sheet (i.e. the holomorphic square root of the bulk
Liouville potential) becomes precisely one of the normalizable massless
scalars in the hyper multiplet. The fact that we add such a term to the
world-sheet action in order to turn up the value 
of $\hat{J}$ is consistent with the fact that the
corresponding field acquires a vacuum expectation value.
More precisely, we have the relations \cite{Hosomichi:2004ph}:
\begin{eqnarray}
\label{mua}
\mu_A &=& \sqrt{\frac{2 k \mu }{\pi}} \sin \pi (-\hat{J}-\hat{m})
\nonumber \\
\bar{\mu}_A &=&  \sqrt{\frac{2 k \mu }{\pi}} \sin \pi (-\hat{J}+\hat{m})
\end{eqnarray}
for the coefficients $\mu_A$ and $\bar{\mu}_A$ of the F-terms on the boundary
of the world-sheet which are holomorphic square roots of the bulk Liouville
potential. In a covariant formulation, and for a zero momentum $k_\mu=0$
ground state, we note that the combination $G^{N=1,nc} |0,0,[\hat{J}] \rangle$
is pure gauge.  We conclude that for very small non-zero $\hat{J}$ and $\hat
m$, 
the null deformation, which will not change the spectrum on the non-compact
brane will correspond to the deformation $\mu_A = - \bar{\mu}_A$ (since the
brane parameter $\hat{m}$ does not change the spectrum). Therefore
the $\hat{J}$ deformation near $\hat{J}=0$ corresponds to the deformation
$\mu_A = + \bar{\mu}_A$. Thus, infinitesimally near $\hat{J}=0=\hat m$, we
have a 
complex scalar field which has one component which becomes the longitudinal
component of the massive vector boson, and an orthogonal component which
acquires a vacuum expectation value.

\subsubsection*{Remark}
An interesting open problem is to attempt to give a mass to the hyper
multiplet while keeping the vector multiplet massless. A naive attempt
to separate the identity brane from the rest of the $\hat{J}=0$ brane
would fail because the remainder does not pass the Cardy
check. Another way to put this is to note that one would generate two
massive hyper multiplets, and minus one massless hyper multiplet.

\subsection{The gauge theory on the compact branes}
We turn to the gauge theory on the compact identity branes. We will be brief
since the theory has been analyzed before. The identity branes are stretched
between two NS5-branes.
The effective gauge theory
below the curvature scale $M_{gap}$ is four dimensional. The low-energy theory
contains one massless vector multiplet of ${\cal N}=2$ supersymmetry in four
dimensions, 
and gives rise to a pure $U(N)$ super-Yang-Mills theory for $N$ compact branes.
They have been discussed as exact boundary states in 
\cite{Eguchi:2003ik}\cite{Israel:2005fn}\cite{Ashok:2005py}.
\subsection{Combining branes}
If we consider simultaneously non-compact and compact branes, we must consider
whether they are mutually supersymmetric. Once we fix the parameters $\hat m$
and $\hat n$ of the
identity brane (which fix its orientation) 
then we must fix the non-compact brane orientation
such that the brane is parallel.  We then preserve eight supercharges again.
At energy scales below the mass gap but above the energy $\hat{J}M_{gap}$, we
have a massive vector multiplet of mass squared $-2 \hat{J} (2\hat{J}-1)/k
\alpha'$, a complex massive vector with a mass squared $-\hat{J}(\hat{J}-1)/k
\alpha'$ as well as the massless vector multiplet on the identity brane. The
real massive vector multiplet and the massless vector multiplet on the
identity brane have been discussed before.

The complex massive vector multiplet embodies a Higss mechanism of the second
type \cite{Fayet:1978ig} in which one vector multiplet eats a scalar of another vector
multiplet and vice versa. In more modern language, we go to a Coulomb 
branch
of
an ${\cal N}=2$ gauge theory. Again, to render the ${\cal N}=2$
gauge theory interactions manifest, we 
would need to analyze in detail the open
string interactions.

\section{Two NS5-branes}
\label{two}
The case of two NS5-branes is an interesting special case. When we spread two
branes on a circle, they are in fact spread on a line (which we will take to
be the $x^6$ direction). We therefore preserve a
larger symmetry group in this case, which includes an $SO(3)_{789}$ isometry 
group (enlarged to
$O(3)$ due to the $Z_2$ operation interchanging the two NS5-branes).
We will use this extension of the symmetry group to make an interesting
 consistency check on our claims in 
previous sections. In the
exact conformal field theory at $k=2$, the minimal model becomes trivial.
The bulk theory and its symmetries were thoroughly analyzed in
\cite{Murthy:2003es}. We recall that at
 level $2$ we can fermionize the asymptotic
angular variable $\theta$ in Liouville theory by the equivalence 
$e^{\pm i \theta} = \frac{1}{\sqrt{2}} (\psi_1 \pm i \psi_2)$. This allows us
to define the $SU(2)_2$ affine currents:
\begin{eqnarray}
J_i &=& (\psi_\rho \psi_i - \frac{1}{2} \epsilon_{ijk} \psi_j \psi_k)
 e^{-\rho},
\end{eqnarray}
where $\psi_3$ is the fermionic partner of the angular variable $\theta$ and
$\psi_\rho$ is the superpartner of the asymptotic radial direction
$\rho$. From the detailed arguments in \cite{Murthy:2003es}, one concludes that the bulk
Liouville deformation can be chosen such as to preserve a diagonal $SU(2)_2$
current algebra. It is this local symmetry algebra on the world-sheet that
generates the $SO(3)_{789}$ isometry group in space-time (in the doubly scaled regime).

We wish to add non-compact branes to this configuration. In doing so we
observe that there is a single non-compact D4-brane that we can add to the
asymptotically flat configuration which will preserve the $SO(3)_{789}$
isometry group. It is the non-compact D4-brane that stretches along the line
separating the $NS5$-branes (namely the $x^6$ direction) and coinciding with
the NS5-branes in the transverse space $x^{7,8,9}$. From the arguments in the
previous section, we expect that that brane should correspond to the
non-compact brane at $\hat{J}=0=\hat{m}$
 after descending down the throat.

We can test this directly in the full exact conformal field theory, using the
symmetry algebra above.  From the results of \cite{Murthy:2003es}\footnote{See
  the bottom of page 14 of \cite{Murthy:2003es}.} 
and \cite{Hosomichi:2004ph}\footnote{See the bottom of page 34 of
  \cite{Hosomichi:2004ph}.} 
it is clear that the
three currents evaluated on the boundary 
correspond precisely to the two marginal
 F-term deformations and the marginal D-term deformation
of boundary $N=2$ Liouville theory. Thus, the boundary deformations form a
triplet of $SO(3)_{789}$. Consequently, when any of these boundary
deformations is turned on, we break $SO(3)_{789}$. We already saw that the
F-term boundary deformations are generically zero at $\hat{J}=0=\hat{m}$ (see
equation (\ref{mua})). So, to check whether $SO(3)_{789}$ is preserved we need
to check whether also the third deformation parameter is zero. It has the 
value \cite{Hosomichi:2004ph}:
\begin{eqnarray}
\tilde{\mu}_A &=& - \nu^{\frac{1}{2}} \frac{\Gamma(-1/k)}{2 \pi k}
 \cos \frac{\pi}{k} (2\hat{J}-1).
\end{eqnarray}
We observe that precisely 
at the value $k=2$, and $\hat{J}=0$ the third boundary deformation
parameter $\tilde{\mu}_A$
 is zero as well. In summary, we used a particularly symmetric set-up 
to provide a powerful consistency check on our identification
of the brane $\hat{J}=0=\hat{m}$ as the brane that splits on two NS5-branes.

\section{Conclusions}
\label{conclusions}

In this paper, we have demonstrated character identities for the $N=2$
superconformal algebra, which lead to the introduction of addition
relations for boundary states. We then showed the appearance of
localized bound states in the spectrum of non-compact branes, both via
an analytic and an algebraic approach.

These interesting properties of non-rational $N=2$ Liouville theory
find applications in string theory. We analyzed one application in
detail, for branes in the background of NS5-branes spread on a
circle. There we showed that the addition relations are reflected in
semi-classical properties of D-branes with three directions transverse
to the NS5-branes. We also gave an exact conformal field theory
analysis of the spectrum on these branes, and showed the appearance of
interesting open string bound states localized deep down the throat of
NS5-branes (in a triple scaling limit).  We performed a preliminary
analysis of the gauge theory interactions that govern these bound
states at low energy.

There remain many interesting open questions related to our work. We mention
only a few interesting routes.
The four-dimensional gauge theory we identified deserves more
investigation. The computation of the interactions of the bound states
at low energy would be a useful enterprise. It would also be
interesting to apply what we have learned about $N=2$ Liouville theory
to configurations that preserve only four supercharges. 
Another direction would be the construction of boundary states that do not
factorize in terms of the 
coset conformal field theories, and find their
interpretation in terms of D-branes in NS5-brane backgrounds. This is
presumably related
to the problem of giving a mass to the hyper multiplet living on the
non-compact brane, while keeping the vector multiplet massless.
We hope to return to some of these open problems in the near future.

\section*{Acknowledgments} 
Our work was supported in part
by the EU under the contract MRTN-CT-2004-005104. We would like to thank
Pierre Fayet for a useful discussion.

\appendix

\section{Full Cardy states in $N=2$ Liouville}
\label{discretecouplings}
In this appendix we review the detailed construction of the full Cardy states
in $N=2$ Liouville theory \cite{Eguchi:2003ik}, including the couplings to the
discrete Ishibashi states. We also compute the annulus amplitudes for
arbitrary discrete branes. We show that discrete characters appear in
the open string channel with 
multiplicities that are generically not positive integers.

We carefully normalize our amplitudes and one-point functions as
follows.  We fix the normalization of the continuous Ishibashi states
such that their overlap is $\delta (P-P')$. Our convention for the
normalization of the closed string momentum (and therefore of the
$\delta$-function) is such that $J=1/2+iP$ determines the Casimir of
an $SL(2,\mathbb{R})$ representation.  We take the $\delta$-function
to be the $\delta$-function on the real half-line, namely
$\int_0^\infty dP \delta(P-P') f(P) = f(P')$ for a function $f$
defined on the positive half-line. Also, we integrate over Ishibashi
states with positive momentum $P$ only. That fixes the normalization
of the one-point functions. In the closed string channel, we will sum
over all integer momenta $2m$ (modulo $2k$).

The continuous and discrete Ishibashi states are defined as\footnote{We consider only the Ishibashi states associated to extended
  characters that appear in the modular transformations of the
  identity character \cite{Eguchi:2003ik}. In the bulk theory, closed
  strings states belongs precisely to these representations
  \cite{Dijkgraaf:1991ba}\cite{Hanany:2002ev}.}:
\be e^{-\pi \frac{c}{3} \frac{z^2}{T}}
 {}_c\langle \langle P,m | \mathrm{prop} |
P',m' \rangle \rangle_c = \delta (P-P')
 \delta_{m,m'} Ch_c^{closed}(P,m;iT,z) \ee
\be e^{-\pi \frac{c}{3} \frac{z^2}{T}} 
{}_d\langle \langle J,r | \mathrm{prop} |
J',r' \rangle \rangle_d = \delta_{J,J'}
 \delta_{r,r'} Ch_d^{closed}(J,m;iT,z) \ee
\be e^{-\pi \frac{c}{3} \frac{z^2}{T}}
 {}_c\langle \langle P,m | \mathrm{prop} | J,r
\rangle \rangle_d = 0 \ee
The closed string propagator is:
\be\label{prop} \mathrm{prop} = e^{\pi \frac{c}{3} \frac{z^2}{T} e^{-\pi T
    H^{(c)}}} 
  e^{i\pi z (J_0+\bar{J}_0)} \ee
where $H^{(c)},J_0,\bar{J}_0$ 
represent the cylinder Hamiltonian and the $U(1)R$ charge operators.

In section \ref{boundary} we introduced the continuous, discrete and identity
branes in $N=2$ Liouville theory, and gave their one-point functions to the
continuous operators. The 
couplings to the discrete operators are:
\be \psi^c_{\hat P,\hat m}(J,r) = 0 \ee
\be \psi^d_{\hat J, \hat r}(J,r) = \frac{i}{\sqrt{2k}} \frac{1}{\sqrt{\sin(\frac{\pi(2J-1)}{k})}}
  e^{-4\pi i \frac{(\hat J+\hat r)(J+r)-(\hat J-1/2)(J-1/2)}{k}} \ee
\be \psi^{\mathbb{I}}_{\hat m}(J,r) = \sqrt{\frac{2}{k}} \sqrt{\sin \left(
    \frac{\pi (2J-1)}{k}\right)} e^{-4\pi i \frac{2\hat m (J+r)}{2k}}.
\ee

\subsection*{Modular transformations} 
\label{modular}
In the computation of annulus amplitudes, modular S-transformations are needed
to switch between the closed string channel and the open string channel. We
give below the relevant modular transformations for the characters of the
extended $N=2$ superconformal algebra.

\be Ch_c^{closed}(P,m;-\frac{1}{\tau},\frac{z}{\tau}) = e^{i\pi \frac{c}{3}
\frac{z^2}{\tau}} \frac{2}{k} \sum_{2m'=0}^{2k-1}e^{-4\pi i
\frac{mm'}{k}} \int dP' \cos\left( 4\pi \frac{P P'}{k} \right) Ch_c^{open}(P',m';\tau,z) \ee

\beq && Ch_d^{closed}(J,r;-\frac{1}{\tau},\frac{z}{\tau}) = e^{i\pi \hat{c}
  \frac{z^2}{\tau}} \times \nonumber \\
&&  \left[ \frac{1}{k} \sum_{2m'=0}^{2k-1}e^{-4\pi i
\frac{(J+r)m'}{k}} \int dP' \frac{\cosh(2\pi P' \frac{k+1-2J}{k}) + e^{i\pi 2 m'}
\cosh(2\pi P' \frac{2J-1}{k})} {2|\cosh(\pi(P'+im'))|^2} Ch_c^{open}(P',m';\tau,z) \right.\nonumber \\
&& + \frac{i}{k} \sum_{r'=0}^{k-1} \sum_{2J'=2}^{k} e^{-4\pi i
  \frac{(J+r)(J'+r')-(J-\frac{1}{2})(J'-\frac{1}{2})}{k}} Ch_d^{open}(J',r';\tau,z)
\nonumber \\ 
&& \left. +\frac{i}{2k} \sum_{r'=0}^{k-1} e^{-4\pi i
  \frac{(J+r)(r'+\frac{1}{2})}{k}}\left(
Ch_d^{open}\left( \frac{1}{2},r';\tau,z \right) +Ch_d^{open}\left(
  \frac{k+1}{2},r';\tau,z \right) \right)\right] \eeq

\beq Ch_{\mathbb{I}}^{closed}(m;-\frac{1}{\tau},\frac{z}{\tau}) &=& e^{i\pi \hat{c}
\frac{z^2}{\tau}} \times
\nonumber \\
& &  \left[ \frac{1}{k} \sum_{2m'=0}^{2k-1}e^{-4\pi i
\frac{mm'}{k}} \int dP' \frac{\sinh(2\pi \frac{P'}{k}) \sinh(2\pi
P')}{|\cosh(\pi(P'+im'))|^2} Ch_c^{open}(P',m';\tau,z) \right. \nonumber \\
&&  + \left. \frac{2}{k} \sum_{r'=0}^{k-1} \sum_{2J'=2}^{k} \sin(\frac{\pi(2J'-1)}{k}) e^{-4\pi i
  \frac{(m(J'+r')}{k}} Ch_d^{open}(J',r';\tau,z) \right] \eeq
In the bulk of the paper, and in the following computations, we leave the
argument $\tau,z$ of the characters implicit but indicate whether we
consider the closed or the open string channel.

\subsection*{Annulus amplitudes}
We compute the annulus amplitude for open strings stretching between two
branes, taking into account the coupling to discrete operators.
\begin{itemize}
\item Identity / any
\be {}_{\mathbb{I}}\langle \hat m'| \mathrm{prop} |\hat m \rangle_{\mathbb{I}} = Ch_{\mathbb{I}}^{open} (\hat m-\hat m') \ee

\be  {}_{\mathbb{I}}\langle \hat m'| \mathrm{prop} |\hat P,\hat m \rangle_c = Ch_c^{open} (\hat P,\hat m-\hat m') \ee

\be  {}_{\mathbb{I}}\langle \hat m'| \mathrm{prop} |\hat J,\hat r \rangle_d = Ch_d^{open} (\hat J,\hat r-\hat m') \ee

\item Continuous / continuous

This overlap is computed in detail in section \ref{bound}. The result is given
by equations (\ref{Zc0}) and (\ref{Zca}).
\item Continuous / discrete
\beq && {}_c\langle \hat P_1,\hat m_1| \mathrm{prop} |\hat J_2\hat ,r_2 \rangle_d =
\int_0^{\infty} dP \left[ \rho_1^{c-d}(P|\hat P_1,\hat J_2)
  Ch_c^{open} (P,\hat J_2+\hat r_2-\hat m1) \right. \nonumber \\
&& \left. + \rho_2^{c-d}(P|\hat P_1,\hat J_2) Ch_c^{open} (P,\hat J_2+\hat r_2-k-\hat m1) \right] \eeq

with 
\be \rho_1^{c-d}(P|\hat P_1,\hat J_2)=\frac{2}{k} \int_0^\infty dP'
\frac{\cos(4\pi\frac{PP'}{k})}{\sinh(2\pi \frac{P'}{k}) \sinh(2\pi P')}
\sum_{\epsilon=\pm 1} \cosh \left( 2\pi P' \left( 1+ \frac{1-2\hat J_2}{k}+i\epsilon
\frac{2\hat  P_1}{k} \right) \right) \ee

\be \rho_2^{c-d}(P|\hat P_1,\hat J_2)=\frac{2}{k} \int_0^\infty dP'
\frac{\cos(4\pi\frac{PP'}{k})}{\sinh(2\pi \frac{P'}{k}) \sinh(2\pi P')}
\sum_{\epsilon=\pm 1} \cosh \left( 2\pi P' \left( \frac{2\hat J_2-1}{k}+i\epsilon
\frac{2\hat  P_1}{k} \right) \right) \ee

\item Discrete / discrete
\beq && {}_d \langle \hat J_1,\hat r_1| \mathrm{prop} |\hat J_2,\hat r_2 \rangle_d =
\int_0^{\infty} dP \sum_{2m=0}^{2k-1}  \rho_1^{d-d}(P,m|\hat J_1,\hat r_1,\hat J_2,\hat r_2)
  Ch_c^{open} (P,m) \nonumber \\
&& + \sum_{r=0}^{k-1} \sum_{2J=2}^{k} \rho_2^{d-d}(J,r|\hat J_1,\hat r_1,\hat J_2,\hat r_2) Ch_d^{open} (J,r)  \nonumber \\
&& +  \sum_{r=0}^{k-1}  \frac{1}{2} \tilde{\rho}_2^{d-d} (\frac{1}{2} ,r|\hat J_1,\hat r_1,\hat J_2,\hat r_2) (Ch_d^{open}
  (\frac{1}{2},r)+ Ch_d(\frac{k+1}{2},r)  \eeq
with
\beq && \rho^{d-d}_1(P,m|\hat J_1,\hat r_1,\hat J_2,\hat r_2)= \int_0^\infty dP' \sum_{2m'=0}^{2k-1}
\frac{\sqrt{2}}{k^{\frac{5}{2}}}\frac{\cos(4\pi\frac{PP'}{k}) }{\sinh(2\pi
  \frac{P'}{k}) \sinh(2\pi P')} e^{4\pi i\frac{m'(-m+(\hat J_1+\hat r_1)-(\hat J_2+\hat r_2))}{k}} \nonumber \\
&& 
\frac{ 1  }{
4|\cosh(\pi(P'+im'))|^2} \times \nonumber \\
& & 
\left( \cosh ( 2\pi P' \frac{k+1-2\hat J_2}{k} ) + e^{2\pi i
      m'}\cosh ( 2\pi P' \frac{2\hat J_2-1}{k} ) \right)  \times
\nonumber \\
& & 
\left( \cosh
    ( 2\pi P' \frac{k+1-2\hat J_1}{k} ) + e^{-2\pi i
    m'}\cosh ( 2\pi P' \frac{2\hat J_1-1}{k} ) \right)   
    \nonumber \\
&&  + \sum_{2J'=2}^{k} \sum_{r'=0}^{k-1} \frac{1}{2k^2} e^{4\pi i
  \frac{(J'+r')(\hat J_1+\hat r_1-(\hat J_2+\hat r_2)-m)-(J'-\frac{1}{2})(\hat J_1-\hat J_2)}{k}}  \frac{\cosh(2\pi P \frac{k+2-4J'}{k}) \cosh(\pi P')}
{2|\cosh(\pi(P+im))|^2 \sin(\pi \frac{2J'-1}{k})} 
 \eeq

\be \rho_2^{d-d}(J,r|\hat J_1,\hat r_1,\hat J_2,\hat r_2)= \frac{i}{2k^2} \sum_{r'=0}^{k-1}
  \sum_{2J'=2}^{k} \frac{ e^{4\pi i
  \frac{(J'+r')(\hat J_1+\hat r_1-(\hat J_2+\hat r_2)-(J+r))+(J'-\frac{1}{2})(J-\frac{1}{2}+\hat J_2-\hat J_1)}{k}}
  } {\sin(\pi
  \frac{2J'-1}{k})} 
 \ee
The spectral density  $\rho_2^{d-d}$ gives the multiplicity of the discrete
characters. The Cardy condition requires that these multiplicities be positive
integers. We can rewrite $\rho_2^{d-d}$ as:
\be \rho_2^{d-d}(J,r|\hat J_1,\hat r_1,\hat J_2,\hat r_2)
= \frac{i}{2k} e^{-\frac{2\pi i}{k} (\hat J_2+\hat r_2-\hat J_1-\hat r_1+J+r)}
 \delta(\hat J_2+\hat r_2-\hat J_1-\hat r_1+J+r \in \frac{k}{2} \mathbb{Z}) 
G(\hat r_2-\hat r_1+r)
\ee
where we introduced the function $G$:
\be G(N) = \sum_{p =1}^{k-1}\frac{e^{-\frac{2 \pi i}{k} p(N+1/2)}}{\sin(\pi
    \frac{p}{k})} \ee
We now prove by induction that:
\be G(N) = -i(k-1)+2i(N[mod\ k]) \ee
where $0 \le N[mod\ k] < k $. 
First, $G(0)=-i(k-1)$. 
Then we write:
\be e^{-\frac{2 \pi i}{k} p(N+1/2)} = -2i  e^{-\frac{2 \pi i}{k} p N}
\sin(\pi \frac{p}{k}) + e^{-\frac{2 \pi i}{k}p (N-1/2)} \ee
so that we can rewrite $G(N)$ in terms of $G(N-1)$:
\beq G(N)&=&-2i\sum_{p=1}^{k-1} e^{-\frac{2 \pi i}{k} p N} + G(N-1) \nonumber
\\
&=& G(N-1)+2i-2ik\, \delta(N \in k \mathbb{Z}) \eeq
This concludes the proof.

Eventually we can write the spectral density $\rho_2^{d-d}$ as:
\beq  \rho_2^{d-d}(J,r|\hat J_1,\hat r_1,\hat J_2,\hat r_2)&=& e^{-\frac{2\pi i}{k} (\hat J_2+\hat r_2-\hat J_1-\hat r_1+J+r)}
 \delta(\hat J_2+\hat r_2-\hat J_1-\hat r_1+J+r \in \frac{k}{2} \mathbb{Z}) \nonumber \\
&&
\left( \frac{k-1-2\left[ (\hat r_2-\hat r_1+r) [ mod\ k] \right] }{2k} \right) \eeq
This function is generically not a positive integer. This invalidates the Cardy
condition for discrete branes. 
\end{itemize}

\subsection*{The self-overlap of the continuous brane $|\hat J=0,\hat r \rangle_c$}
The continuous brane $|\hat J=0,\hat r \rangle_c$ can be written as
the sum of an identity and two discrete branes, according to the
addition relation (\ref{c=i+d+d}). In section \ref{bound} we computed
the self-overlap of the continuous brane $|\hat J=0,\hat r
\rangle_c$. We now 
consider the self-overlap of the sum of branes on the right-hand side
of (\ref{c=i+d+d}):

\beq && \left( {}_{\mathbb{I}} \langle \hat r | + {}_d \langle \hat J= k/2, \hat r
    | + {}_d \langle \hat J= 1, \hat r-1 | \right)  \mathrm{prop}  \left( |
  \hat r \langle_{\mathbb{I}} + | \hat J = k/2, \hat r \rangle_d +  | \hat J =
  1, \hat r-1 \rangle_d \right) \nonumber \\
&& = Ch_{\mathbb{I}}^{open}(0) + 2 Ch_d^{open}(k/2,0) + 2 Ch_d^{open}(1,0)
\nonumber \\
&& +
\left(  {}_{d} \langle \hat J = k/2, \hat r
    | + {}_d \langle \hat J =1, \hat r -1 | \right)  \mathrm{prop}  \left( 
   | \hat J = k/2, \hat r \rangle_d +  | \hat J =
  1, \hat r-1 \rangle_d \right)
 \eeq
Let's compute the contribution to the discrete characters of the overlap of the discrete branes:
\beq && {}_{d} \langle \hat J = k/2, \hat r|   \mathrm{prop} 
   | \hat J = k/2, \hat r \rangle_d 
= {}_d \langle \hat J =1, \hat r -1 |   \mathrm{prop}  
  | \hat J = 1, \hat r-1 \rangle_d \nonumber \\
&& = \sum_{2J=2}^{k} \left( \frac{k-1-2[-J
[mod\ k]]}{2k} Ch_d^{open}(J,-J) -\frac{k-1-2[k/2 - J [mod\ k]]}{2k} Ch_d^{open}(J,k/2-J) \right) \nonumber \\
&&+\ continuous
\eeq
\beq && {}_{d} \langle \hat J = k/2, \hat r |  \mathrm{prop}   
   | \hat J =  1, \hat r-1 \rangle_d \nonumber \\
 && = \sum_{2J=2}^{k} \left(
  -\frac{k-1-2\left\{ [-J[mod\ k]]-1[mod\ k] \right\} }{2k} Ch_d^{open}(J,-J) \right. \nonumber \\
 &&+
 \left. \frac{k-1-2[k/2 - J -1 [mod\ k]]}{2k} Ch_d^{open}(J,k/2-J) \right) \nonumber \\
&&+\ continuous
\eeq
\beq && {}_d \langle \hat J =1, \hat r -1 |  \mathrm{prop} 
   | \hat J = k/2, \hat r \rangle_d \nonumber \\
 && = \sum_{2J=2}^{k} \left(
  -\frac{k-1-2\left\{ [-J[mod\ k]]+1[ mod\ k]\right\} }{2k} Ch_d^{open}(J,-J) \right. \nonumber \\
 &&+\left.
  \frac{k-1-2[k/2 - J +1 [mod\ k]]}{2k} Ch_d^{open}(J,k/2-J) \right) \nonumber \\
&&+\ continuous
\eeq
Which gives (for $k \neq 1$) :
\beq && \left(  {}_{d} \langle \hat J = k/2, \hat r
    | + {}_d \langle \hat J =1, \hat r -1 | \right)  \mathrm{prop}  \left( 
   | \hat J = k/2, \hat r \rangle_d +  | \hat J =
  1, \hat r-1 \rangle_d \right)  \nonumber \\
&& = - Ch_d^{open}(1,-1) -
Ch_d^{open}(k/2,0) + continuous \eeq
Thus we recover the result (\ref{ZzeroJ}), with the correct multiplicity 
one for the discrete characters:
\be\label{ZzeroJbis} \langle P=i/2,M=0|P=i/2,M=0 \rangle = Ch_d^{open}(1,-1) + Ch_d^{open}(k/2,0) +
Ch_{\mathbb{I}}^{open}(0) + continuous
\ee
In the special case $k=1$, the overlap of discrete branes contains
only continuous representations.  As a consequence the discrete
characters in the overlap of the continuous brane (\ref{ZzeroJbis})
appear with multiplicity two. This is also what we found in section
\ref{bound} of the bulk of the paper.

\section{The example of $k=3$ NS5-branes}
\label{k=3}
To make the formulas for the annulus partition function for open strings ending
on a continuous $\hat{J}=0$ brane on one side and an identity brane on the other
side more concrete, we
discuss one example in detail (beyond the case of $k=2$ NS5-branes treated in
more detail in \cite{Ashok:2005py}). We concentrate on $k=3$ Neveu-Schwarz five-branes.
To get to grips with the ingredients of the partition function, 
we need to compute the relevant superparafermionic characters. They are given
by \cite{Gepner:1987qi}:
\begin{eqnarray}
\chi(j,n,s;\tau,z) &=& \sum_{p \bmod k-2} c^{2j}_{n+4p-s} (\tau)
\Theta_{2n+(4p-s)k,2(k-2)k} (\tau, \frac{2 (k-2)z}{k})
\end{eqnarray}
and the character is zero when $2j+n+s \neq 0\ [\bmod 2]$.
For $k=3$ NS5-branes we have a bosonic $SU(2)$ level
$k_{bos}=1$. Since we also have the equivalence relation
$(j,n,s) \equiv (\frac{k_{bos}}{2}-j,s+2,n+k)$, we can restrict
ourselves to characters with $j=0$. Using identities between string
functions \cite{Gepner:1986hr}
 that implies that we will only need the string
function $c^0_0$ (defined in \cite{Kac:1984mq}). 
Moreover, we sum over $p$ modulo one, so we can set $p=0$.
We find the identity:
\begin{eqnarray}
\chi(j,n,s;\tau,z) &=&  c^{0}_{n-s} (\tau)
\Theta_{2n-3s,6} (\tau, z^{\frac{1}{3}})
\end{eqnarray}
where $n=s \, (\bmod \, 2)$ to have a non-zero result. 
We are left with twelve
non-trivial supercoset characters that can be labeled by $n-3s \, ( \bmod 12)$.
We find the conformal weights and R-charges for the primaries
in these representations of the superconformal algebra at $c=3-\frac{6}{3}=1$
(see also \cite{Maldacena:2001ky} appendix F):
\begin{eqnarray}
\begin{array}{cc}
h& Q_{MM} \\
 0 & 0  \\
1/24 & \pm 1/6  \\
1/6  &  \pm 1/3 \\
3/8 &  \pm 1/2 \\ 
2/3 &  \pm 2/3 \\ 
25/24 & \pm 5/6 \\
3/2 &  \pm 1
\end{array}
\end{eqnarray}
Note that in the case of highest conformal weight, we have two degenerate
ground states, which explains the fact that we have thirteen entries in the table.
We recall that the minimal model $U(1)_R$ charge is given by the formula 
$Q_{MM} = \frac{s}{2} - \frac{n}{k}$.
To compute the partition function, we compute the separate terms in the sum:
\begin{eqnarray}
Z_{(k=3)} &=& \frac{1}{2} (Z^{NS} - Z^{\tilde{NS}}) - \frac{1}{2} (Z^R \mp
Z^{\tilde{R}}).
\end{eqnarray}
Let's concentrate on the first term:
\begin{eqnarray}
Z^{NS} &=& \frac{\theta_3^2(\tau)}{\eta(q)^6}
\times
\nonumber \\
& & 
((\chi(0,0,0) + \chi(0,0,2)) Ch_{c}^{NS} (J=0,2m=0)
\nonumber \\
& & 
+(\chi(0,2,0) + \chi(0,2,2)) Ch_{c}^{NS} (J=0,2m=4)
\nonumber \\
& & 
+(\chi(0,4,0) + \chi(0,4,2)) Ch_{c}^{NS} (J=0,2m=2)).
\end{eqnarray}
The factors corresponding to the minimal model and the non-compact model will
contain the following states. The first term contains a ground state
at $h-c/24=-1/2$. At $h-c/24=-1/4$ the first term 
also contains one state each of $U(1)_R$ charge $\pm 1$. The second term
contains one
such state with $U(1)_R$ charge
$-1$ and the third term one state with $U(1)_R$ charge $+1$. 
When combined with the flat space factor:
\begin{eqnarray}
\frac{\theta_3^2(\tau)}{\eta(q)^6} &=&  q^{-1/4} + (2z^{-1}+2z) q^{1/4} +
\dots
\nonumber 
\end{eqnarray}
we get on the one hand the $4$ bosonic states with charges $\pm 1$ from
the identity in the non-trivial factors, and the second term for the flat space
factors. This is the vector multiplet. On the other hand, we can take the
trivial factor in flat space-time, and we get $2$ states from the excited
levels of the first character as well as two states from the ground states of
the second and third term.
\begin{eqnarray}
&=& q^{-1/2} + (4z^{-1} + 4z) q^{0} + (6z^{-2}+24+6z^2) q^{\frac{1}{2}} + 
(4 z^{-3} + 60 z^{-1} + 60 z + 4 z^3) q^1 +  \dots
\nonumber
\end{eqnarray}
After GSO projection, we will be left with only the first term, which gives
rise to eight massless bosonic states. 
These correspond to the primaries $\psi_{-1/2}^{2,3,4,5}$ acting on the ground
state in the non-trivial factors, as well as the following four states (indicated
by their $U(1)_R$ charges and dimension in the first and second factor conformal field theory,
 in the second and third line):
\begin{eqnarray}
G_{-1/2}^{\pm, nc} |0> & & 
\nonumber \\
(+1/6,-1/3) \otimes (+1/3,-2/3)
\nonumber \\
(+1/6,+1/3) \otimes (+1/3,+2/3)
\end{eqnarray}
These are the massless states that we identified in the annulus
partition function in the bulk of the paper
(when we put $k=3$ in those formulas). To complete
the exercise, we quote the
 Ramond sector partition function (suppressing the $z$-dependence):
\begin{eqnarray}
Z^R &=& 16 q^0 + 256 q^1 + \dots
\nonumber 
\end{eqnarray}
which precisely cancels the NS contribution. The cancellation is a particular
application of a large class of identities between superconformal characters
that were proven on the basis of the spectral flow argument of Gepner
adapted to the open string channel in an appendix to \cite{Ashok:2005py}.

  \end{document}